\documentclass[useAMS,usenatbib,10pt]{mn2e}
\usepackage{times}  % to get nice fonts for PDF
\usepackage{listings}
\usepackage{graphics}
\usepackage{subfigure}
\usepackage{amssymb}
\usepackage{hyperref}
\usepackage{aas_macros}
\usepackage{lscape}
\usepackage{rotating}
\usepackage[export]{adjustbox}
\usepackage{graphicx}

\def\eagle{{\sc eagle}}
\def\simprop{ \lower .75ex \hbox{$\sim$} \llap{\raise .27ex \hbox{$\propto$}} }

\title[Angular momentum of galaxies]{Angular momentum evolution of galaxies in EAGLE}
\author[Claudia del P. Lagos et al.]{
\parbox[t]{\textwidth}{
\vspace{-1.0cm}
Claudia del P. Lagos$^{1,2,3}$\thanks{E-mail: claudia.lagos@icrar.org}, Tom Theuns$^{4}$, Adam R. H. Stevens$^5$, Luca Cortese$^1$, Nelson D. Padilla$^{6,7}$, Timothy A. Davis$^{8}$, Sergio Contreras$^{6}$, Darren Croton$^{5}$}
\vspace*{6pt} \\
$^{1}$International Centre for Radio Astronomy Research (ICRAR), M468, University of Western Australia, 35 Stirling Hwy, Crawley, WA 6009, Australia.\\
$^{2}$Australian Research Council Centre of Excellence for All-sky Astrophysics (CAASTRO), 44 Rosehill Street Redfern, NSW 2016, Australia.\\
$^{3}$Kavli Institute for Theoretical Physics,  Kohn Hall, University of California, Santa Barbara, CA 93106, United States.\\
$^{4}$Institute for Computational Cosmology, Department of Physics,
University of Durham, South Road, Durham, DH1 3LE, UK.\\
$^{5}$Centre for Astrophysics \& Supercomputing, Swinburne University of Technology, Hawthorn, VIC 3122, Australia.\\
$^{6}$Instituto de Astrof\'sica, Pontificia Universidad Cat\'lica de Chile, Avda. Vicu\~na Mackenna 4860, 782-0436 Macul, Santiago, Chile.\\
$^{7}$Centro de Astro-Ingenier\'ia, Pontificia Universidad Cat\'olica de Chile, Avda. Vicu\~na Mackenna 4860, 782-0436 Macul, Santiago, Chile.\\
$^{8}$Astronomy, Cardiff University, Queens Buildings, The Parade, Cardiff CF24 3AA, United Kingdom.
\vspace*{-0.5cm}}

\begin{document}

%\date{Accepted ???. Received ???; in original form ???}

\pagerange{\pageref{firstpage}--\pageref{lastpage}} \pubyear{2016}

\maketitle

\label{firstpage}

\begin{abstract}
We use the \eagle\ cosmological hydrodynamic simulation suite to study the specific angular momentum of galaxies, $j$, with the aims of (i) investigating the physical causes behind the wide range of $j$ at fixed mass and (ii) examining whether simple, theoretical models can explain the seemingly complex and non-linear nature of the evolution of $j$. We find that $j$ of the stars, $j_{\rm stars}$, and baryons, $j_{\rm bar}$, are strongly correlated with stellar and baryon mass, respectively, with the scatter being highly correlated with morphological proxies such as gas fraction, stellar concentration, (u-r) intrinsic colour, stellar age and the ratio of circular velocity to velocity dispersion. We compare with available observations at $z=0$ and find excellent agreement. We find that $j_{\rm bar}$ follows the theoretical expectation of an isothermal collapsing halo under conservation of specific angular momentum to within $\approx 50$\%, while the subsample of rotation-supported galaxies are equally well described by a simple model in which the disk angular momentum is just enough to maintain marginally stable disks. We extracted evolutionary tracks of the stellar spin parameter of \eagle\ galaxies and found that the fate of their $j_{\rm stars}$ at $z=0$ depends sensitively on their star formation and merger histories. From these tracks, we identified two distinct physical channels behind low $j_{\rm stars}$ galaxies at $z=0$: (i) galaxy mergers, and (ii) early star formation quenching. The latter can produce galaxies with low $j_{\rm stars}$ and early-type morphologies even in the absence of mergers.
\end{abstract}


\section{Introduction}

The formation of galaxies can be a highly non-linear process, with many physical mechanisms interacting 
simultaneously (see reviews by \citealt{Baugh06,Benson10b}). Notwithstanding all that potential complexity, 
early studies of galaxy formation stressed the importance of three quantities 
to describe galaxies: mass, $M$, angular momentum, $J$, and energy, $E$ \citep{Peebles69,Doroshkevich70,Fall80,White84}; 
or alternatively, one can define the specific angular momentum, $j\equiv J/M$, which 
contains information on the scale length and rotational velocity of systems. It is therefore 
intuitive to expect the relation between $j$ and $M$ to contain fundamental information.% of galaxies.
 
Studies such as \citet{Fall80}, \citet{White91}, \citet{Catelan96a} and \citet{Mo98}, showed that 
many properties of galaxies, such as flat rotation curves, and the Tully-Fisher relation 
could be obtained in the Cold Dark Matter (CDM) framework if $j$ of baryons is similar to that of the halo and 
is conserved in the process of disk formation (although 
conservation does not need to be strict, but within a factor of $\approx 2$; \citealt{Fall83}). 
The situation is of course different for the mass and 
energy of galaxies, which can vary significantly throughout their evolution due to accretion, star formation and 
dissipative processes, such as galaxy mergers.
Theoretical models of how $j$ of halos evolves in a CDM universe predict $j\propto \lambda\,M^{2/3}$, 
where $\lambda$ is the spin parameter of the halo (e.g. \citealt{White84}; \citealt{Catelan96a}; \citealt{Mo98}). 
If $j$ of baryons is conserved throughout the formation of galaxies, then a similar relation should apply to galaxies.
These models generally assume that halos collapse as their spherical overdensity 
reaches a threshold value, and in that sense neglect mergers.
Due to the dissipative nature of the latter, one would expect significant 
changes in the relation between $j$ and $M$ of halos and galaxies \citep{Zavala08,Sales12,Romanowsky12}. 

Hydrodynamic simulations used to suffer from catastrophic loss of angular momentum, producing 
galaxies that were too compact and too low $j$ compared to observations \citep{Steinmetz99,Navarro00}.
This problem was solved by improving the spatial resolution and including efficient 
feedback 
(e.g. \citealt{Kaufmann07}; \citealt{Zavala08}; \citealt{Governato10}; \citealt{Guedes11}; \citealt{Danovich15}). 
A new generation of simulations have 
 immensely improved in spatial resolution, volume and sophistication of the sub-grid physics included, 
allowing the study of angular momentum loss in galaxies statistically.
For example, simulations such as \eagle\ \citep{Schaye14}, Illustris \citep{Vogelsberger14} and 
Horizon-AGN \citep{Dubois14} achieve spatial resolutions of $\approx 700\,\rm pc$ (physical units), 
volumes of $(100\,\rm Mpc)^3$, and include models for metal cooling, star formation and 
stellar and active galactic nucleus (AGN) feedback. These simulations contain thousands of galaxies 
with stellar masses $>10^{10}\,\rm M_{\odot}$.

Observationally, \citet{Fall83} presented the first study of the relation between $j$ of the stellar component, $j_{\rm stars}$,  
and stellar mass. \citet{Fall83} found that both spiral and elliptical galaxies follow a relation that is close to $j\propto M^{2/3}$, but with 
spiral galaxies having a normalisation $\approx 5$ times larger than elliptical galaxies. 
Recently, this was extended by \citet{Romanowsky12} and \citet{Fall13} in a sample of $\approx 100$ galaxies. 
These studies confirmed that the power-law index 
of the relation was close to $2/3$ for their entire galaxy population 
and that ellipticals galaxies had significantly lower $j$ than spiral galaxies at a given mass. 

\citet{Obreschkow14b} presented the most accurate measurements of $j$ in the stellar, neutral gas and total baryon components of galaxies 
out to large radii ($\approx 10$ times the disk scale length) in a sample of $16$ late-type galaxies of the HI Nearby Galaxy Survey 
(THINGS; \citealt{Walter08}) 
and found (i) galaxies follow a relation close to $j_{\rm stars} \propto M^{2/3}_{\rm stars}$ and $j_{\rm bar} \propto M^{2/3}_{\rm bar}$, where 
$M_{\rm stars}$, $M_{\rm bar}$ and $j_{\rm bar}$ are the stellar mass, 
baryon mass (stars plus neutral gas) and baryon specific angular momentum respectively, (2) 
the scatter in the $j_{\rm bar}$-$M_{\rm bar}$ and $j_{\rm stars}$-$M_{\rm stars}$ 
relations is strongly correlated with the bulge-to-total stellar mass 
ratio and the neutral gas fraction 
(neutral mass divided by baryon mass; $f_{\rm gas, neutral}$). By 
fixing the bulge-to-total stellar mass ratio, \citet{Obreschkow14b} found that  
$j_{\rm bar} \propto M_{\rm bar}$.
Using the \citet{Toomre64} stability model, surface density of the gas in galaxies and a flat exponential disk, 
\citet{Obreschkow16} found that the atomic gas fraction in galaxies is $\propto (j_{\rm bar}/M_{\rm bar})^{1.12}$.
\citet{Obreschkow14b} argued that 
under the assumption that bulges in spiral galaxies form through disk instabilities, 
one could understand the relation between $j_{\rm stars}$, stellar mass and bulge-to-total stellar mass 
ratio from the model above. 
\citet{Stevens16}, using a semi-analytic model, 
showed that disk instabilities play a major role in regulating the $j_{\rm stars}-M_{\rm stars}$ sequence for spiral galaxies, 
consistent with the picture of \citet{Obreschkow14b}. 

To measure $j$ accurately in galaxies, requires spatially resolved kinematic information.
The pioneering work of the SAURON \citep{Bacon01} and ATLAS$^{\rm 3D}$ \citep{Cappellari11} surveys, on samples of galaxies that   
comprised $260$ early-type galaxies in total,  
showed that the stellar kinematics and distributions of stars are not strongly correlated, and thus morphology 
is not necessarily a good indicator of the dynamics of galaxies \citep{Krajnovic13}. 
Based on these surveys, \citet{Emsellem07,Emsellem11} coined the terms {\it slow} and {\it fast} rotators, and proposed 
 the $\lambda_{\rm R}$ parameter, which measures how rotationally or dispersion-dominated a galaxy is and is closely connected to $j_{\rm stars}$, 
as a new, improved scheme to classify galaxies. 
\citet{Naab14} showed later that such a classification is also applicable for galaxies in 
hydrodynamic simulations.
Unfortunately, accurate measurements of $j$ have only been presented for a few hundred galaxies.
The future, however, is bright: the advent of integral field spectroscopy (IFS) and the new generation of radio and millimeter telescopes 
promises a revolution in the field. 

Currently, the Sydney-AAO Multi-object Integral field spectrograph (SAMI; \citealt{Croom12}) survey is observing 
$\approx 3,200$ galaxies for which resolved kinematics will be available \citep{Bryant15}. 
Similarly, high-resolution radio telescopes, such as the Square Kilometre Array (SKA), promise to collect information that would allow 
the measurement of $j$ for few thousand galaxies 
during its first years \citep{Obreschkow15b}, truly revolutionising our understanding of the build-up of 
angular momentum in galaxies. 
\citet{Cortese16} presented the first measurements of the $j_{\rm stars}$-$M_{\rm stars}$ relation 
for $297$ galaxies in SAMI, and 
found that, for the entire sample and for a relation of the form $j_{\rm stars}\propto M^{\alpha}_{\rm stars}$, 
$\alpha\approx 0.7$, close to the theoretical expectation of $2/3$, 
but when studied in subsamples of different morphological types 
$\alpha$ varies from $0.69$ for elliptical galaxies to $0.97$ for spiral galaxies.
 Cortese et al. found that the dispersion of the $j_{\rm stars}-M_{\rm stars}$ relation 
is correlated with morphological proxies such as S\'ersic index and light concentration.
These new results have not yet been examined in simulations.

In this paper we explore two long-standing open questions of how $j$ evolves in galaxies:
(i) how does $j$ depend with mass, and what are the most relevant secondary galaxy properties, and  
(ii) how well do simple, theoretical models explain the evolution of $j$ in a complex, non-linear 
hydrodynamical simulations. 
In our opinion, \eagle\ is the ideal testbed for this experiment due to 
the spatial resolution achieved, the large volume that allows us to statistically assess these relations and also
the growing amount of evidence that the simulation produces a realistic galaxy population. 
For instance, \eagle\ reproduces well the relations between star formation rate (SFR) and stellar mass
(\citealt{Furlong14}; \citealt{Schaye14}), the colour bi-modality of galaxies 
(\citealt{Trayford15,Trayford16}), the molecular and atomic gas fractions as a function of stellar mass 
(\citealt{Lagos15}; \citealt{Bahe15}; \citealt{Crain16}), 
 and the co-evolution of stellar mass, SFR and gas \citep{Lagos15b}. 

So far, simulations have been used to 
test theoretical models for the evolution of angular momentum. 
For instance, \citet{Zavala15} presented a study of the build-up of angular momentum of the stars, 
cold gas and dark matter in \eagle, and showed that disks form mainly after the {\it turnaround} 
epoch (epoch of maximum expansion of halos, after which 
they collapse into virialised structures, approximately conserving specific angular momentum) 
 while bulges formed before turnaround, explaining why bulges have much lower $j$. 
\citet{Zavala15} also compared the $j_{\rm stars}$-$M_{\rm stars}$ relation for \eagle\ galaxies at $z=0$ with 
the observations of \citet{Romanowsky12} and found general agreement.
\citet{Teklu15} and \citet{Pedrosa15} 
also found that that the positions of galaxies in the $j_{\rm stars}$-$M_{\rm stars}$ relation 
is correlated with the bulge-to-total stellar mass ratio in the Magneticum and Fornax simulations, respectively.
Similarly, \citet{Genel15} presented an analysis of the effect of baryon processes 
on the $j_{\rm stars}$-$M_{\rm stars}$ relation in the Illustris simulation 
and confirmed previous results that feedback is a key process preventing catastrophic angular momentum loss.
Here we investigate several galaxy properties 
that have been theoretically and/or empirically 
proposed to be relevant for the relationship between $j$ and mass in \eagle, 
and extend previous work by exploring a larger parameter space of galaxy properties that could 
determine the positions of galaxies in the $j$-mass relation of 
 different baryonic components of galaxies.
 We also perform the most, to our knowledge, comprehensive comparison between hydrodynamic simulations 
and observations of $j$ to date.

This paper is organised as follows. In $\S$~\ref{EagleSec} we give a
brief overview of the simulation, and 
 describe how the dynamic and kinematic properties of galaxies used in this paper are calculated.
In $\S$~\ref{theoryback} we give a theoretical background that we then use to interpret our 
results.
In $\S$~\ref{jz0sec} we explore the dependence of $j$ on galaxy properties at $z=0$ and 
present a comprehensive comparison with observations.
{In $\S$~\ref{StructureEvolution} we analyse in detail the evolution of $j$ of the different baryonic components of galaxies,  
and identify average evolutionary tracks of $j_{\rm stars}/M^{2/3}_{\rm stars}$.}
Here we also compare the evolution of $j$ in \eagle\ with simple, theoretical 
models to study how closely these models can reproduce the trends seen in \eagle.
We discuss our results and present our conclusions in $\S$~\ref{ConcluSec}.
In Appendix~\ref{ConvTests} we present `weak' and `strong' convergence tests (terms introduced by \citealt{Schaye14}),
and in Appendix~\ref{ScalingRelations} we present additional 
scaling relations between the specific angular momentum of stars and baryons and other galaxy properties.

\section{The EAGLE simulation}\label{EagleSec}

\begin{table}
\begin{center}
  \caption{Features of the Ref-L100N1504 simulation used in this paper. The row list:
    (1) comoving box size, (2) number
    of particles, (3) initial particle masses of gas and (4) dark
    matter, (5) comoving gravitational
    softening length, and (6) maximum physical comoving Plummer-equivalent
    gravitational softening length. Units are indicated in each row. \eagle\
    adopts (5) as the softening length at $z\ge 2.8$, and (6) at $z<2.8$. }\label{TableSimus}
\begin{tabular}{l l l l}
\\[3pt]
\hline
& Property & Units & Value \\
\hline
(1)& $L$ & $[\rm cMpc]$ & $100$\\
(2)& \# particles &  & $2\times 1504^3$ \\
(3)& gas particle mass & $[\rm M_{\odot}]$ & $1.81\times 10^6$\\
(4)& DM particle mass & $[\rm M_{\odot}]$ & $9.7\times 10^6$\\
(5)& Softening length & $[\rm ckpc]$ & $2.66$\\
(6)& max. gravitational softening & $[\rm pkpc]$& $0.7$ \\
\hline
\end{tabular}
\end{center}
\end{table}

The \eagle\ simulation suite\footnote{See {\tt
    \footnotesize http://eagle.strw.leidenuniv.nl} and {\tt
    \footnotesize http://www.eaglesim.org/} for images, movies and data
  products. A database with many of the galaxy properties in \eagle\ is publicly available and described in 
\citet{McAlpine15}.}  (described in detail by \citealt{Schaye14}, hereafter
S15, and \citealt{Crain15}, hereafter C15) consists of a large number of cosmological
hydrodynamic simulations with different resolutions, cosmological volumes and subgrid models,
adopting the \citet{Planck14} cosmological parameters.
S15 introduced a reference model, within which the parameters of the
sub-grid models governing energy feedback from stars and accreting black holes (BHs) were calibrated to ensure a
good match to the $z=0.1$ galaxy stellar mass function and 
the sizes of present-day disk galaxies.

In Table~\ref{TableSimus} we summarise the parameters 
of the simulation used in this work, including the number of
particles, volume, particle masses, and spatial resolution.  
Throughout the text we use pkpc to denote proper kiloparsecs and 
cMpc to denote comoving megaparsecs. 
A major aspect of the \eagle\ project is the use of
state-of-the-art sub-grid models that capture unresolved physics.
The sub-grid physics modules adopted by \eagle\ are: (i) radiative cooling and 
photoheating, (ii) star formation, (iii) stellar evolution and enrichment, 
(iv) stellar feedback, and (v) black hole growth and active galactic nucleus (AGN) feedback 
(see S15 for details on how these are modelled and implemented in \eagle). In addition, 
the fraction of atomic and molecular gas in gas particle is calculated in post-processing 
following \citet{Lagos15}.

The \eagle\ simulations were performed using an extensively modified
version of the parallel $N$-body smoothed particle hydrodynamics (SPH)
code {\sc gadget-3} (\citealt{Springel08}; \citealt{Springel05b}).
Among those modifications are updates to the SPH technique, which are collectively referred to as 
`Anarchy' (see \citealt{Schaller15b} for an analysis of the impact of these changes on 
the properties of simulated galaxies compared to standard SPH). We use {\sc SUBFIND} 
 (\citealt{Springel01}; \citealt{Dolag09}) to identify self-bound overdensities of particles within halos (i.e. substructures). 
These substructures are the galaxies in \eagle. 

\subsection{Calculation of dynamic and kinematic properties of galaxies in EAGLE}\label{StructureProperties}

Here we describe how we measure velocity dispersion of the stars; specific angular momentum and the stellar, neutral gas
{(atomic and molecular gas mass, in both components hydrogen plus helium)} and total baryon components; 
rotational velocity; and $\lambda_{\rm R}$ parameters. We measure these properties 
 in apertures that range from $3$~pkpc to $500$~pkpc in all galaxies with $M_{\rm stars}>10^9\,\rm M_{\odot}$.
We also calculate the half-mass radius of stars, which we use to compute 
$j$ of the stellar component in a physically meaningful aperture, which is also comparable to those used in observations.

We calculate the $1$-dimensional velocity dispersion of the stars perpendicular to the midplane of the disk. 
We do this by calculating the velocity relative to the centre of mass 
$\Delta v_{\rm i}=\mid\vec{v}_{\rm i}-\vec{v}_{\rm COM}\mid$. Here, 
$\vec{v}_{i}$ and $\vec{v}_{\rm COM}$ are the velocity vectors of the $i$-th particle and that of the centre of mass, with 
the latter being calculated using all the particles of the subhalo (DM plus baryons). 
We then take the component of the velocity vector above parallel to the total stellar angular momentum vector 
(i.e. using all the star particles in the sub-halo), $L_{\rm stars}$, and compute:

\begin{equation}
\sigma_{\rm 1D,\star}(r) = \sqrt{\frac{\sum_{\rm i}\,m_{\rm i}\,(\Delta v_{\rm i}\,{\rm cos(\theta_{\rm i})})^2}{\sum_{\rm i}\,m_{\rm i}}}.
\label{Sigma1dstar}
\end{equation}
 
\noindent Here, $\rm cos(\theta_{\rm i}) = \Delta \vec{v}_{\rm i} \cdot \vec{L}_{\rm stars} / \mid\Delta \vec{v}_{\rm i}\mid\, \mid \vec{L}_{\rm stars} \mid$.
We calculate the rotational velocity of a galaxy from the specific angular momentum of the baryons (star and gas particles
with a non-zero neutral gas fraction), $\vec{j}_{\rm bar}$, as:

\begin{equation}
V_{\rm rot}(r) \equiv \frac{\mid \vec{j}_{\rm bar}(r)\mid}{r}.
\label{spin}
\end{equation}

\noindent We do not include ionised gas in the calculation of $\vec{j}_{\rm bar}$ 
because its angular momentum is negligible compared to the stellar and neutral gas components, 
and because it makes it easier to compare with observations, in which this is measured 
from the stars, HI and H$_2$ (e.g. \citealt{Obreschkow14b}). 
We calculate $\vec{j}$ as

\begin{equation}
\vec{j} = \frac{\sum_{\rm i}\,m_{\rm i}\,(\vec{r}_{\rm i}-\vec{r}_{\rm COM}) \times (\vec{v}_{\rm i}-\vec{v}_{\rm COM})}{\sum_{\rm i}\,m_{\rm i}},
\label{spin2}
\end{equation}

\noindent where $\vec{r}_{\rm i}$ and $\vec{r}_{\rm COM}$ are the position vectors (from the origin of the box) 
of particle $i$ and the centre of mass. To calculate $j$ of the stars, neutral gas and baryons, we use star particles only, 
gas particles that have a non-zero neutral gas fraction only, and the latter two types of particles, respectively. 
 
To calculate $\vec{j}(r)$, $\sigma_{\rm 1D}(r)$ and $V_{\rm rot}(r)$, we use particles enclosed in $r$. This way 
we avoid numerical noises due to the small number of particles that could be used if we were instead measuring 
these quantities in annuli.
{However, when we measure the $\lambda_{\rm R}$ parameter, first introduced by
\citet{Emsellem07}, we need to calculate these quantities in annuli 
as defined in \citet{Naab14}.} This parameter measures how rotationally supported a galaxy is:

\begin{equation}
\lambda_{\rm R}(r) = \frac{\sum^{N(r)}_{i=1} m_{\rm \star,i}\,r_{\rm i}\,V_{\rm rot}(r_{\rm i})}{\sum^{N(r)}_{i=1}m_{\rm \star,i}\,r_{\rm i}\,\sqrt{V^2_{\rm rot}(r_{\rm i})+\sigma^2_{\rm 1D,\star}(r_{\rm i})}}.
\label{lambdar}
\end{equation}

\noindent Here, the sum is over all the radial bins from the inner one to $r$, $N(r)$ is the number of radial bins enclosed 
within $r$, and $m_{\rm \star,i}$ is the stellar mass enclosed in each radial bin. 
This means that this quantity depends on the chosen bins. Here we choose bins of $3$~pkpc of width, to be comfortably 
above the resolution limits, but we tested that the higher resolution simulations return similar relations between 
$j_{\rm stars}-M_{\rm stars}-\lambda_{\rm R}$. 
Values of $\lambda_{\rm R}$ close to zero indicate 
dispersion-supported galaxies, while values close to $1$ indicate rotation-dominated galaxies. Typically, in observations 
$\lambda_{\rm R}$ has been measured within an effective radius (that encloses half of the light of a galaxy), and thus we 
use $\lambda_{\rm R}$ measured within a half-mass radius of the stellar component, $r_{\rm 50}$. 
From Eq.~\ref{lambdar}, one would expect a correlation between $j_{\rm bar}$ and 
$\lambda_{\rm R}$, given that $j_{\rm bar}$ appears in the nominator of Eq.~\ref{lambdar}.

Throughout the text we denote the specific angular momentum of stars as $j_{\rm stars}$ and that of the neutral gas as 
$j_{\rm neutral}$, unless otherwise stated, these are calculated with all the particles within $r_{\rm 50}$.
The latter is a $3$-dimensional radius, rather than a projected one. This choice is made to be able to compare with observations, that 
usually measure $j$ within $r_{50}$. {When we use `(tot)', 
for example $j_{\rm stars}(\rm tot)$, we refer to the measurements of $j$ made using all of the 
particles of that class that belong to the sub-halo hosting the galaxy.
In addition and unless otherwise stated, we impose 
$r_{\rm 50}>1$~pkpc (above the spatial resolution of the simulation), to avoid numerical artifacts.}

In Appendix~\ref{ConvTests} we analyse the resolution limits of the simulation used here by comparing with higher 
resolution runs of \eagle, focusing on $j_{\rm stars}$, $j_{\rm neutral}$ and
$j_{\rm bar}$, as a function of stellar mass, neutral gas mass and baryon mass, respectively.
 We place a conservative limit in stellar, neutral gas and baryon mass above which $j_{\rm stars}$, $j_{\rm neutral}$ and
$j_{\rm bar}$ are well converged (either by measuring within $r_{50}$ 
or within a fixed aperture). These limits are $M_{\rm stars}=10^{9.5}\,\rm M_{\odot}$, 
$M_{\rm bar}=10^{9.5}\,\rm M_{\odot}$ and 
$M_{\rm neutral}=10^{8.5}\,\rm M_{\odot}$ for the simulation used here (Table~\ref{TableSimus}). 
Throughout the paper we show results down to stellar and baryon masses of $10^9\,\rm M_{\odot}$, and 
neutral masses of $10^8\,\rm M_{\odot}$, but show these conservative resolution limits to mark roughly when 
the results become less trustworthy. 

Throughout the paper we study trends as a function of stellar,   
neutral gas and baryon mass. Neutral gas corresponds to the atomic plus molecular gas mass, 
while the baryon mass is defined as $M_{\rm bar}\equiv M_{\rm stars}+M_{\rm neutral}$ (here we neglect the ionised gas). 
The latter definition is close to what observations consider to be the baryon mass of galaxies \citep{Obreschkow14b}.
Following S15, all these properties are measured in $3$-dimensional apertures of $30$~pkpc.
The effect of the aperture is minimal as shown by \citet{Lagos15} and S15.
Once these quantities are defined, we calculate the neutral gas fraction as:

\begin{equation}
f_{\rm gas,neutral}\equiv \frac{M_{\rm neutral}}{(M_{\rm neutral}+M_{\rm stars})}.\label{fgasneutral}
\end{equation}

Note that mass measurements are close to {\it total} masses, while $j$ is measured in an aperture which is a function of 
$r_{50}$. We do this because in observations masses are calculated from broadband photometry, in the case of stellar mass, 
and from emission lines, in the case of HI and H$_2$ masses, that enclose the entire galaxy, which means that observations 
recover masses that are close to {\it total} masses. 
However, when $j$ is measured, high quality, resolved kinematics maps are usually required, which are 
in many cases only present for the inner regions of galaxies, such as within $r_{50}$. 

\section{Theoretical background}\label{theoryback}
 
To interpret our findings in \eagle, it is useful to set a theoretical background first, with the expectations of 
simple models for how $j$ evolves in galaxies under given circumstances, such as conservation of specific angular momentum. 
With this in mind, we introduce here the predictions of the isothermal collapsing halo model \citep{White84,Catelan96a,Mo98} 
and of the more recent model of \citet{Obreschkow14b} which connects $j$ with the stability of disks and the grow of bulges.

In the model of an isothermal collapsing halo with negligible angular momentum losses, there is a relation 
 between $j$, mass and spin parameter of the halo, $\lambda$. This relation is given by:

\begin{equation}
j_{\rm h} = \frac{\sqrt{2}\,G^{2/3}}{(10\,H)^{1/3}}\,\lambda\,M^{2/3}_{\rm h},
\label{jhalo}
\end{equation}  

\noindent where $j_{\rm h}$ and $M_{\rm h}$ are the halo specific angular momentum and mass, respectively, 
$G$ is Newton's gravity constant and $H$ is the Hubble parameter \citep{Mo98}.
Under the assumptions of conservation of $j$, one can write $j_{\rm bar}= j_{\rm h}$, and 
we can replace $M_{\rm h}$ by the baryon mass, using the baryon fraction in each halo, 
$M_{\rm bar} = f_{\rm b} \, M_{\rm h}$.
In $\S$~\ref{StructureProperties} we introduced the 
$\lambda_{\rm R}$ parameter, and based 
on \citet{Emsellem07}, we can relate the halo spin with $\lambda_{\rm R}$ as 
$\lambda_{\rm R}\approx 10\,\lambda$ via assuming that galaxies are $\approx 10$ smaller than 
their halo, that $j_{\rm halo}\sim j_{\rm stars}$ and a fixed mass model (so that the relation
between the gravitational and effective radii of galaxies is fixed\footnote{This is a very drastic
simplification, given the wide variety of mass distributions found in galaxies \citep{Jesseit09}. {In addition, \citet{Kravtsov13} shows that
 the $2\sigma$ scatter around that relation of the size of galaxies and their halo is large, i.e. of $\approx 0.5$~dex.}}). 
{\citet{Kravtsov13} found that $r_{50}\approx 0.015\,r_{\rm halo}$, where $r_{\rm halo}$ is the 
halo virial radius. \citet{Obreschkow14b} showed in local spiral galaxies that $j_{\rm stars}$ and 
$j_{\rm bar}$ converge at $r_{\rm g}\approx 5-6\,r_{50}$, 
and since here we care about the total $j$, we take $r_{\rm g}\approx 0.1\,r_{\rm halo}$ as the relevant galaxy size.} 
 Using the approximations above, 
we can rewrite Eq.~\ref{jhalo} in terms of the baryon component

\begin{equation}
j_{\rm bar} \approx \frac{\sqrt{2}\,G^{2/3}\,f^{-2/3}_{\rm b}}{10\,(10\,H)^{1/3}}\,\lambda_{\rm R}\,M^{2/3}_{\rm bar}, 
\label{jbarpred}
\end{equation} 

\noindent which we evaluate as

\begin{equation}
\frac{j_{\rm bar}}{\rm pkpc\,km\,s^{-1}} \approx 4.26\times10^{-5}\,f^{-2/3}_{\rm b}\,\lambda_{\rm R}\,\left(\frac{M_{\rm bar}}{\rm M_{\odot}}\right)^{2/3}.
\label{jhaloiso}
\end{equation}

\noindent Eq.~\ref{jhaloiso} is similar to Eq.~$15$ in \citet{Romanowsky12}, except that here 
we write it in terms of the baryon content and $\lambda_{\rm R}$. If for example we were to assume that $f_{\rm b}$ is constant and equal to the 
Universal baryon fraction measured by \citet{Planck14}, $f_{\rm b}=0.157$, then Eq.~\ref{jhaloiso} becomes,

\begin{equation}
\frac{j_{\rm bar}}{\rm pkpc\,km\,s^{-1}} \approx 1.46\times10^{-4}\,\lambda_{\rm R}\,\left(\frac{M_{\rm bar}}{\rm M_{\odot}}\right)^{2/3}.
\label{jhaloisoexample}
\end{equation}

\noindent In $\S$~\ref{CompToModels} we compare Eqs.~\ref{jhaloiso}~and~\ref{jhaloisoexample} with those of \eagle.

In the model of stability of disks, \citet{Obreschkow16} showed that by assuming 
a flat exponential disk that is marginally stable, {a relationship between $M_{\rm bar}$, $j_{\rm bar}$ and 
the atomic gas fraction of galaxies is reached}:

\begin{equation} 
f_{\rm atom}\equiv \frac{M_{\rm atom}}{(M_{\rm neutral}+M_{\rm stars})}={\rm min}\left[1,2.5\,\left(\frac{j_{\rm bar}\,\sigma_{\rm gas}}{G\,M_{\rm bar}}\right)^{1.12}\right].
\label{stabilitymodel2}
\end{equation}

\noindent {Here, $M_{\rm atom}$ is the atomic gas mass (hydrogen plus helium) and $\sigma_{\rm gas}$ is the 
velocity dispersion of the gas in the interstellar medium of galaxies. In this model 
$f_{\rm atom}$ is a good predictor of $j_{\rm bar}$ in galaxies, but saturates in gas-rich systems. 
\citet{Obreschkow16} showed that local, isolated galaxies follow this relation very closely.}

We therefore study $j$ as a function of mass, neutral gas fraction and 
spin parameter. In addition, previous studies by \citet{Fall83}, \citet{Romanowsky12}, \citet{Fall13} 
 argued that the morphology of galaxies is a key parameter correlated with the positions of galaxies in the 
$j$-mass plane, so we also study $j$ as a function of several morphological indicators, such as stellar concentration, 
central stellar surface density, optical colour and stellar age. The latter have been connected to morphology 
and quenching of star formation by several observational and theoretical 
works (e.g. \citealt{Shen03,Linttot08,Bernardi10,Woo15,Trayford16}). We define the stellar concentration as 
the ratio between the radii containing $90$\% and $50$\% of the stellar mass, $r_{90}/r_{50}$.
The latter is close to the observational definition which uses the Petrosian radii in the SDSS $r$-band 
containing $50$\% and $90$\% of the light (e.g. \citealt{Kelvin12}). In the case of the central stellar surface density, $\mu_{\rm stars}$, 
observers have used the value within $1$~pkpc \citep{Woo15}. However, since the resolution of \eagle\ is very close to 
that value, we decide to choose a slightly larger aperture of $3$~pkpc to measure $\mu_{\rm stars}$. Unless otherwise stated, 
 $\mu_{\rm stars}$ is always measured within the inner $3$~pkpc of galaxies. 
We study $j$ as a function of the intrinsic (u-r) colours of galaxies, 
$\rm (u^{*}-r^{*})$, and the mass-weighted stellar ages, $\langle\rm age_{\rm stars}\rangle$. 
The latter properties were taken from the \eagle\ public database, described in \citet{McAlpine15}.

\section{The specific angular momentum of galaxies in the local universe}\label{jz0sec}

\begin{figure*}
\begin{center}
\includegraphics[width=0.95\textwidth]{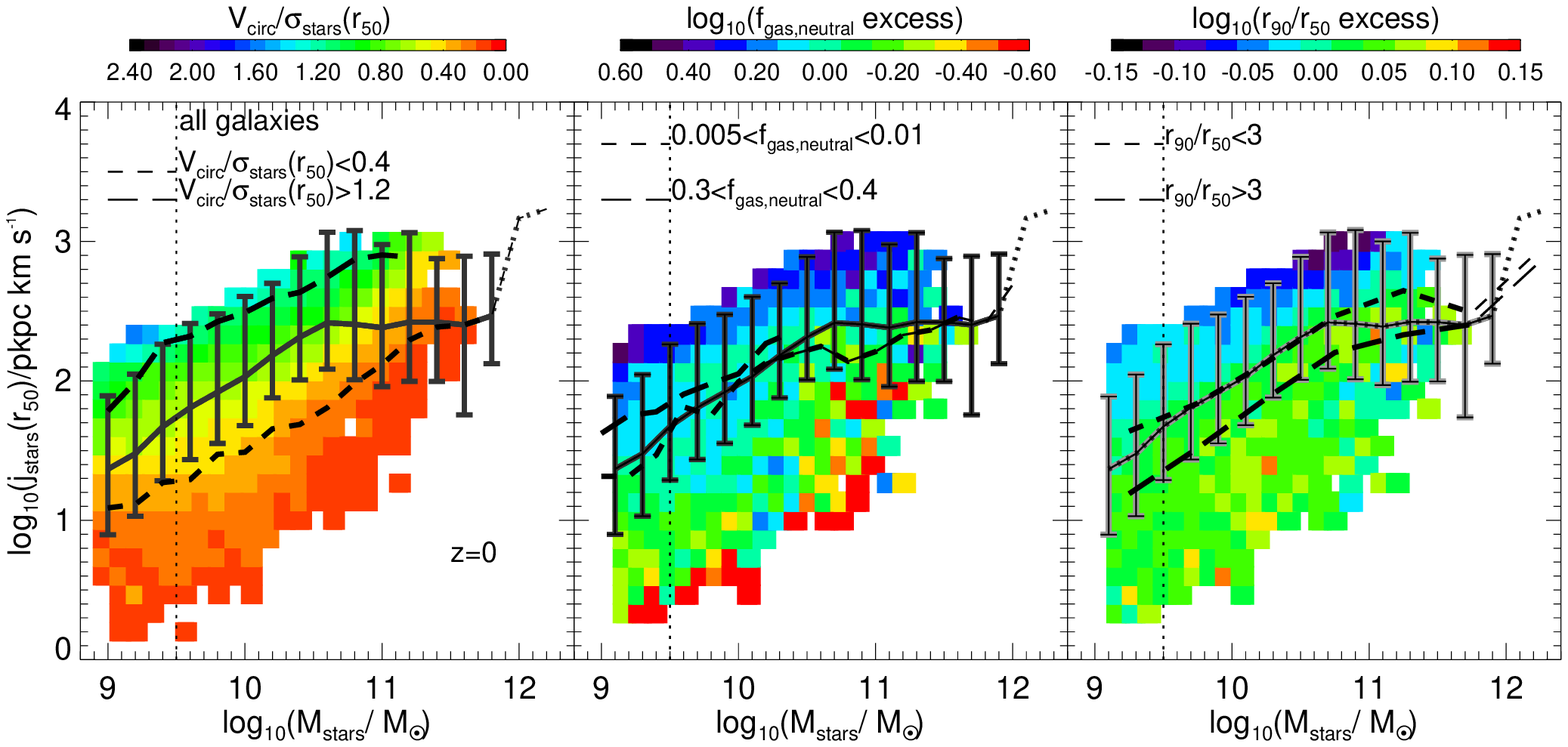}
\includegraphics[width=0.95\textwidth]{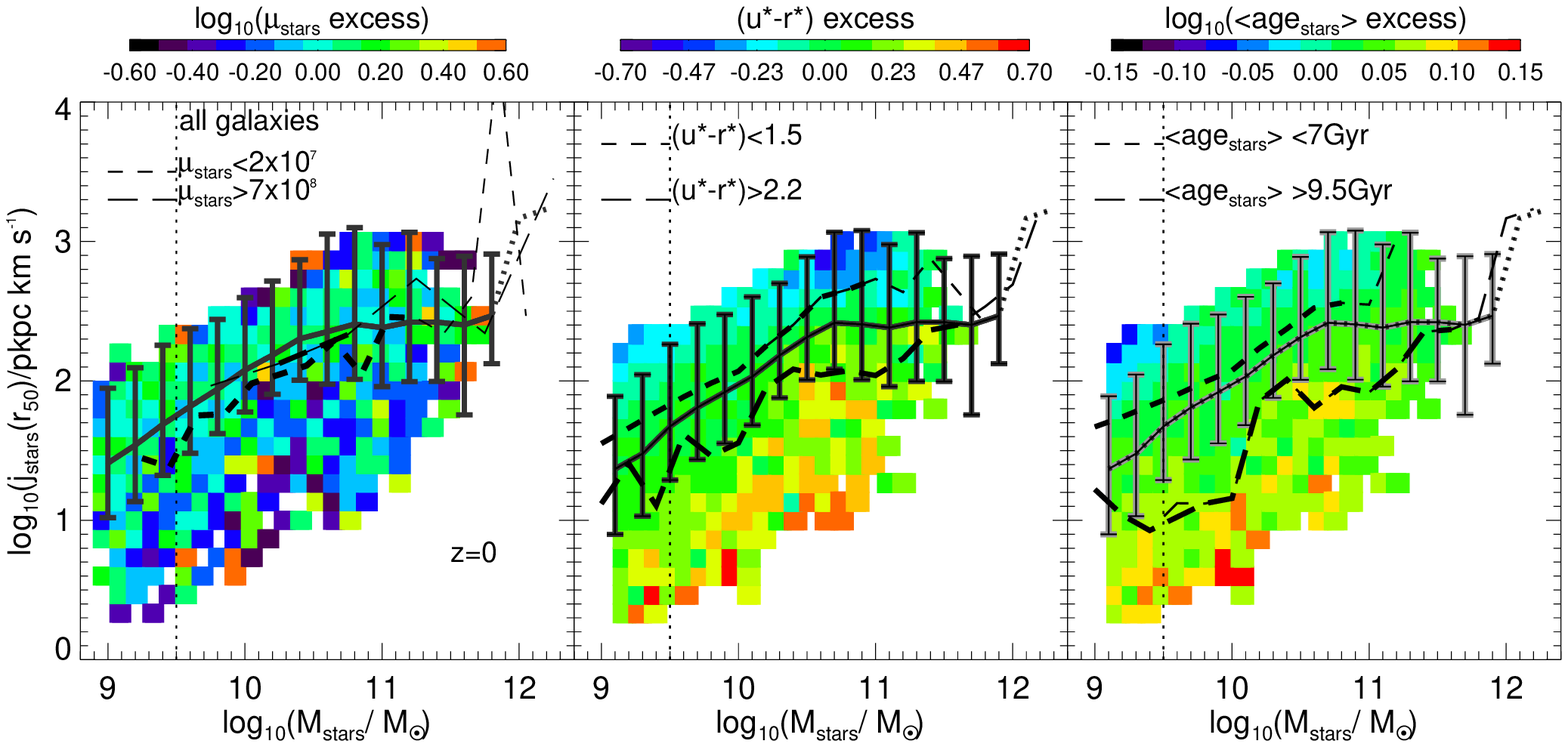}
\caption{The specific angular momentum of the stars, measured with all the particles within 
$r_{50}$, as a function of stellar mass at $z=0$ for all galaxies with $M_{\rm stars}>10^9\,\rm M_{\odot}$. 
In each panel, the $j_{\rm stars}(r_{50})-M_{\rm stars}$ plane is colour coded according to 
the median $V_{\rm rot}(r_{50})/\sigma_{\rm stars}(r_{50})$ (top-left panel), 
neutral gas fraction (top-middle panel), $r_{\rm 50}/r_{90}$ (top-right panel), 
$\mu_{\rm stars}$ (measured within the inner $3$~pkpc; bottom-left panel), 
(u$^*$-r$^*$) colour (bottom-middle panel) and mass-weighted stellar age, $\langle\rm age_{\rm stars}\rangle$ (bottom-right panel), 
in pixels with $\ge 5$ objects. 
Here excess is defined as the ratio between the median in the 2-dimensional bin divided by the median 
at fixed stellar mass, so that negative (positive) values indicate galaxies to be below (above) the median at fixed stellar mass. 
In each panel the solid line and error bars indicate the median and 
$16^{\rm th}$ to $84^{\rm th}$ percentile range of $j_{\rm stars}(r_{50})$ at fixed stellar mass, while 
the short and long-dashed lines show two subsamples of galaxies (as labelled in each panel).
 Bins with $<10$ galaxies are shown as thinner lines.
For reference, the vertical dotted line shows a conservative stellar mass limit above which $j_{\rm stars}$ 
is well converged for the resolution of the simulation.} 
\label{JMEAGLE}
\end{center}
\end{figure*}

\begin{figure}
\begin{center}
\includegraphics[width=0.49\textwidth]{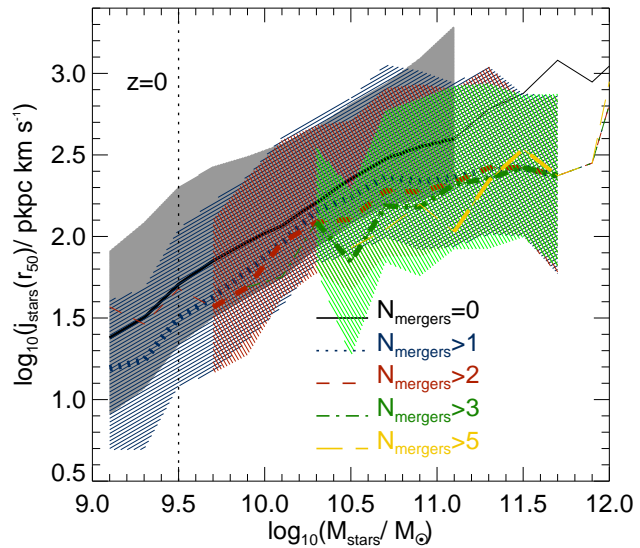}
\caption{$j_{\rm stars}(r_{\rm 50})$ {as a function of stellar mass, 
 at $z=0$, for galaxies with $M_{\rm stars}>10^9\,\rm M_{\odot}$ in \eagle.
We show this relation for galaxies selected by the number of mergers
they suffered throughout their history, as labelled. Lines show the median relations, while the
shaded regions show the $16^{\rm th}-84^{\rm th}$ percentile range, but only for the cases of
$N_{\rm mergers}=0,\,>1,\,>2,\,>3$.
For reference, the vertical line shows a conservative stellar mass 
limit above which $j_{\rm stars}$ is well converged for the resolution of the simulation
(see Appendix~\ref{ConvTests} for details). Note that here we only consider as galaxy mergers those with a baryonic mass ratio
$\ge 0.1$. Mass ratios below that are considered to be below the resolution limit \citep{Crain16}.}}
\label{JsJbEvoMergers}
\end{center}
\end{figure}

\begin{figure}
\begin{center}
\includegraphics[width=0.49\textwidth]{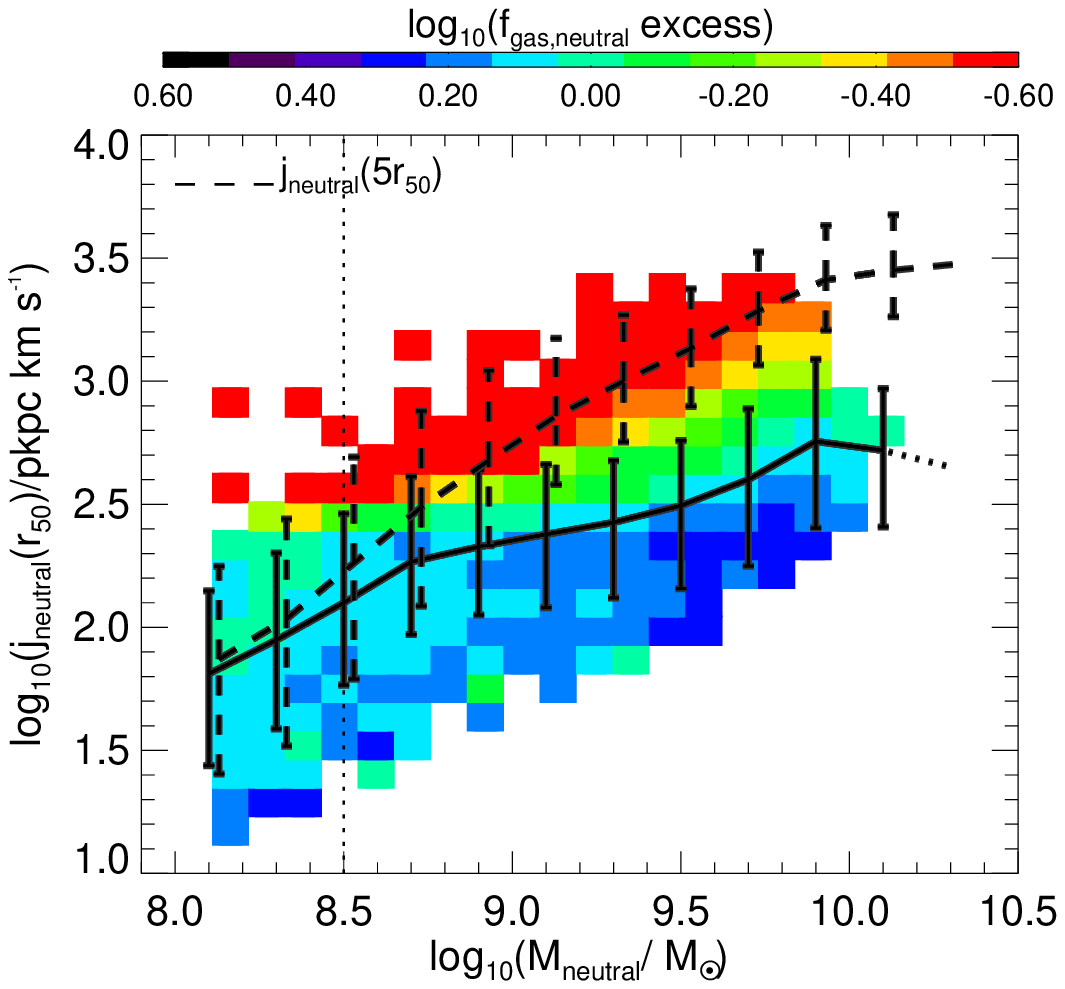}
\caption{$j_{\rm neutral}$, measured within $r_{\rm 50}$, as a function of the neutral gas mass $z=0$ for galaxies with $M_{\rm stars}>10^9\,\rm M_{\odot}$. The solid line with error bars indicate the median and
$16^{\rm th}$-$84^{\rm th}$ percentile range, respectively. 
Pixels with more than $5$ galaxies are coloured by the normalised median 
$f_{\rm gas,neutral}$, as indicated by the colour bar at the top. 
The dashed line with error bars show the median and $16^{\rm th}$-$84^{\rm th}$ percentile range, respectively, 
of the relation between $j_{\rm neutral}$, measured within $5\times r_{\rm 50}$, and  the neutral gas mass.}
\label{JMNEAGLEz0}
\end{center}
\end{figure}

The top left panel of Fig.~\ref{JMEAGLE} shows the correlation between $j_{\rm stars}(r_{50})$ and $M_{\rm stars}$ for galaxies with 
$M_{\rm stars}>10^9\,\rm M_{\odot}$ at $z=0$ in \eagle. We find a moderately tight correlation between 
$j_{\rm stars}$ and $M_{\rm stars}$, with a scatter (i.e. standard deviation) of $\approx 0.6$~dex at fixed stellar mass. 
Galaxies with $10^9\,\rm M_{\odot}<M_{\rm stars}\lesssim 10^{10.6}\,\rm M_{\odot}$ display an increasing $j_{\rm stars}$ 
with increasing $M_{\rm stars}$, while for higher stellar mass galaxies, $j_{\rm stars}$ flattens. This is related to the 
transition from disk-dominated to bulge-dominated galaxies in \eagle\ at $z=0$ (\citealt{Zavala15}) and 
to the occurrence of galaxy mergers. {The latter is shown in Fig.~\ref{JsJbEvoMergers}, which shows
the $j_{\rm stars}-M_{\rm stars}$ relation for galaxies that have had no mergers, 
at least one merger, and successively up to at least $5$ mergers.
We identified mergers using the merger trees available in the \eagle\ database \citep{McAlpine15}. 
Here we do not distinguish mergers that took place recently or far in the past,
but just count their occurrence.
At fixed stellar mass, galaxies with a higher incidence of mergers have significantly lower $j_{\rm stars}$. For example, at
$M_{\rm stars}\approx 10^{10.7}\,\rm M_{\odot}$, galaxies that had never had a merger have $0.5$~dex higher
$j_{\rm stars}$ than galaxies that suffered more than $5$ mergers in their lifetime. We present a comprehensive analysis of the effect of mergers
on $j_{\rm stars}$ in an upcoming paper (Lagos et al. in preparation).}
 
In \eagle\ we find that several galaxy properties that trace morphology are related to 
$\lambda_{\rm R}$ and $j_{\rm stars}$ (Fig.~\ref{Morphology}), and 
 thus the scatter of the $j_{\rm stars}-M_{\rm stars}$ relation is also expected to correlate with these properties.
Indeed we find clear trends with all these properties in the middle and right panels of 
Fig.~\ref{JMEAGLE}. 
To remove the trend between these properties and stellar mass, we coloured pixels by the 
median value of each property in each pixel divided by the median in the stellar mass bin. We name 
this ratio as {\it excess}. Galaxies with lower $f_{\rm gas,neutral}$, redder optical colours and 
higher stellar concentrations have lower $j_{\rm stars}$.
We do not find a relation between the scatter in the $j_{\rm stars}-M_{\rm stars}$ relation with $\mu_{\rm stars}$ 
This is interesting, as recently \citet{Woo15} suggested that $\mu_{\rm stars}$ 
is a good proxy of morphology. This is not seen in \eagle\ as there 
is very little correlation between being rotationally- or dispersion-dominated and $\mu_{\rm stars}$. 
We cannot rule out at this point that the lack of correlation could be due to $\mu_{\rm stars}$ being measured 
here in apertures that are much larger than what observers use ($3$~vs.~$1$~pkpc).

We find that {the scatter of the $j_{\rm stars}-M_{\rm stars}$ relation is most strongly correlated with 
the $V_{\rm rot}/\sigma_{\rm stars}$ ratio and the gas fraction excess} (top left and middle panels of Fig.~\ref{JMEAGLE}).
{The trend with $V_{\rm rot}/\sigma_{\rm stars}$ is obtained almost by construction, given that 
$V_{\rm rot}\propto j_{\rm bar}$ and at $z=0$ $j_{\rm stars}\sim j_{\rm bar}$ due to the low gas fractions 
most galaxies have (note that the latter is not necessarily true for very gas-rich galaxies).} 
Galaxies with $\rm log_{10}(f_{\rm gas, neutral}\, excess)<-0.5$ 
have $\approx 1.5$~dex lower $j_{\rm stars}$ than those with $\rm log_{10}(f_{\rm gas, neutral}\, excess)>0.3$, at fixed stellar mass. 
{We do not find any differences between central and satellite galaxies, which is not necessarily 
surprising given that the angular momentum of the stars 
follows the angular momentum of the inner DM halo, rather than the total halo \citep{Zavala15}, and thus it is less likely to be strongly affected 
by galaxies becoming satellites and any associated stripping of their outer halo.}

We find that \eagle\ galaxies with large values of $r_{90}/r_{50}$ have lower $j_{\rm stars}$ 
at fixed stellar mass (see for example the short- and long-dashed lines in the right panel
of Fig.~\ref{JMEAGLE}). If we instead measure 
$j_{\rm stars}$ out to $5$ times $r_{50}$, the relation between the scatter of the 
$j_{\rm stars}-M_{\rm stars}$ relation and $r_{90}/r_{50}$ 
mostly disappears (not shown here), indicating that this correlation arises only if we look at the central parts of galaxies. 
As for the intrinsic $\rm (u^*-r^*)$ colour, we find that red galaxies, $\rm (u^*-r^*)>2.2$, have 
$0.3$~dex lower $j_{\rm stars}$ than their bluer counterparts, $\rm (u^*-r^*)<1.5$, at fixed stellar mass (bottom middle panel 
of Fig.~\ref{JMEAGLE}). 
A similar difference is found between galaxies that have mass-weighted stellar ages $\rm \langle\rm age_{\rm stars}\rangle>9.5$~Gyr, 
and their younger counterparts with $\rm \langle\rm age_{\rm stars}\rangle<7$~Gyr, at fixed stellar mass.

A major conclusion that can be drawn from Fig.~\ref{JMEAGLE} is that \eagle\ reproduces the 
observational trends of late-type galaxies having much larger $j_{\rm stars}$ than 
early-type galaxies \citep{Fall83,Romanowsky12,Fall13}. This is seen in most of the morphological indicators we use.
In addition, \citet{Zavala15} showed that this trend is also obtained in \eagle\ using the distribution of circular orbits as a proxy for 
morphology. In Fig.~\ref{Morphology} we show the correlation between the morphological indicators used here and the 
$\lambda_{\rm R}$ parameter, which is widely used in the literature to define slow and fast rotators.

{Very similar correlations to those shown in Fig.~\ref{JMEAGLE} are find in the $j_{\rm bar}-M_{\rm bar}$ plane (shown in 
Fig.~\ref{JBAREAGLE}). The most important difference is that we do not find a strong correlation between the scatter in the 
$j_{\rm bar}-M_{\rm bar}$ relation and $r_{90}/r_{50}$.}

Interestingly, in the $j_{\rm neutral}-M_{\rm neutral}$ relation,  
Fig.~\ref{JMNEAGLEz0} shows that galaxies with high $f_{\rm gas,neutral}$ lie {\it below} the median. 
This trend remains when we study 
$j_{\rm neutral}$ out to larger radii.
{We interpret this trend as due to two factors: (i) as gas is consumed in star formation, galaxies move to the left of the diagram, 
and (ii) stars preferentially form from low-$j_{\rm neutral}$ gas, so by taking some of this low $j$ out, 
the $j_{\rm neutral}$ of the remaining gas increases, 
and hence galaxies also move up on the diagram.}

\subsection{Comparisons to observations}
\begin{figure}
\begin{center}
\includegraphics[width=0.48\textwidth]{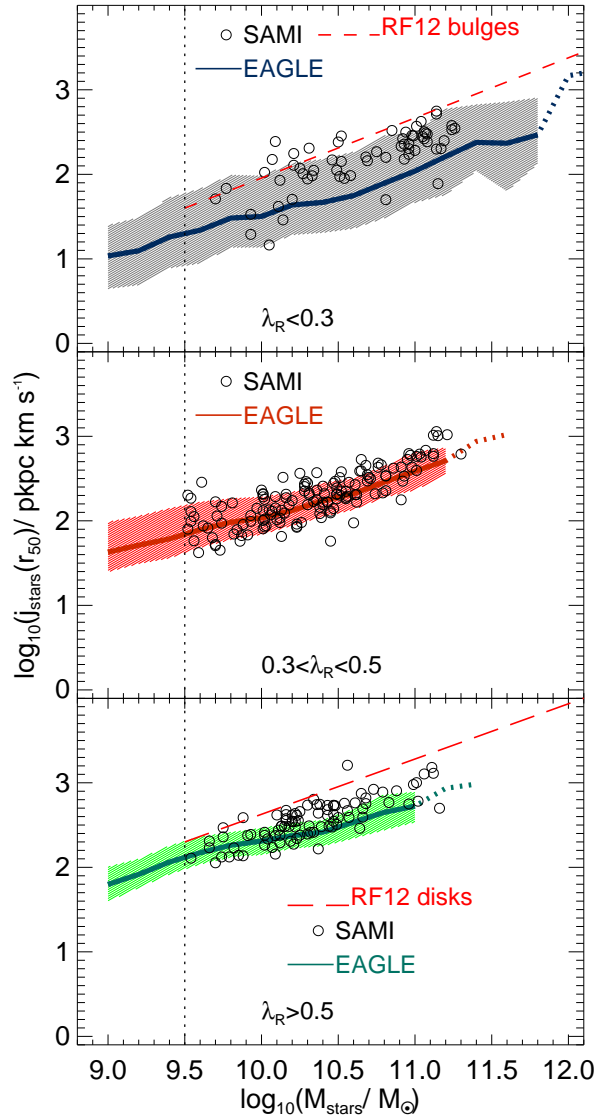}
\caption{The $j_{\rm stars}(r_{\rm 50})-M_{\rm stars}$ relation at $z=0$ in three bins of $\lambda_{\rm R}$, 
as labelled in each panel. Lines and shaded regions show the 
median and $16^{\rm th}$ to $84^{\rm th}$ percentile ranges, respectively. 
Bins with less than $10$ objects are shown as dotted lines.
{Symbols show the observations of the SAMI survey \citep{Cortese16}, and are shown in the different panels 
for the same bins of $\lambda_{\rm R}$ as adopted in \eagle.}
We also show the observational result of \citet{Romanowsky12} (RF12) for bulges in the top panel (short-dashed line), and 
disks in the bottom panel (long-dashed line) (Fig.~$14$ in \citealt{Romanowsky12}).
For reference, the vertical dotted lines show a conservative stellar mass 
limit above which $j_{\rm stars}$ is well converged for the resolution of the simulation.
We find a very good agreement with the measurements of SAMI.}
\label{JMcomp}
\end{center}
\end{figure}

\begin{figure}
\begin{center}
\includegraphics[width=0.49\textwidth]{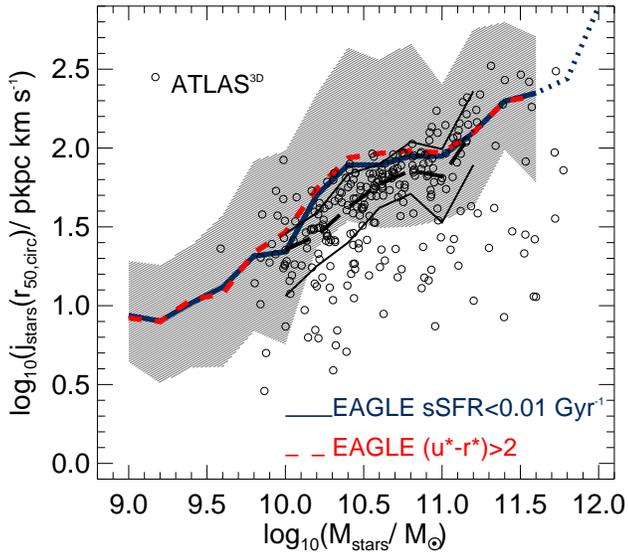}
\caption{$j_{\rm stars}$ as a function of stellar mass for galaxies in \eagle\ at $z=0$ with 
$M_{\rm stars}>10^9\,\rm M_{\odot}$ and with $\rm sSFR<0.01\,\rm Gyr^{-1}$ 
(solid line with shaded region) or with $\rm (u^*-r^*)>2$ (dashed line).
The selections in sSFR and $\rm (u^*-r^*)$ colour are chosen to select passive objects in \eagle\, which 
is an effective way of selecting early-type galaxies.
Here, $j_{\rm stars}$ in \eagle\ is measured within the circularised half-mass radius, 
$r_{\rm 50,circ}$. 
The lines show medians and the shaded region show the 
median and $16^{\rm th}$ to $84^{\rm th}$ percentile range for the sSFR-selected sample. 
The scatter for the colour-selected sample is very similar and thus for clarity is not shown here.
Circles show individual ATLAS$^{\rm 3D}$ observations, while the long-dashed and thin solid lines show the median 
and the $16^{\rm th}$ to $84^{\rm th}$ percentile range of these observations in bins with $\ge 10$ galaxies.}
\label{JMcompA3D}
\end{center}
\end{figure}

\begin{figure*}
\begin{center}
\includegraphics[width=0.355\textwidth,valign=t]{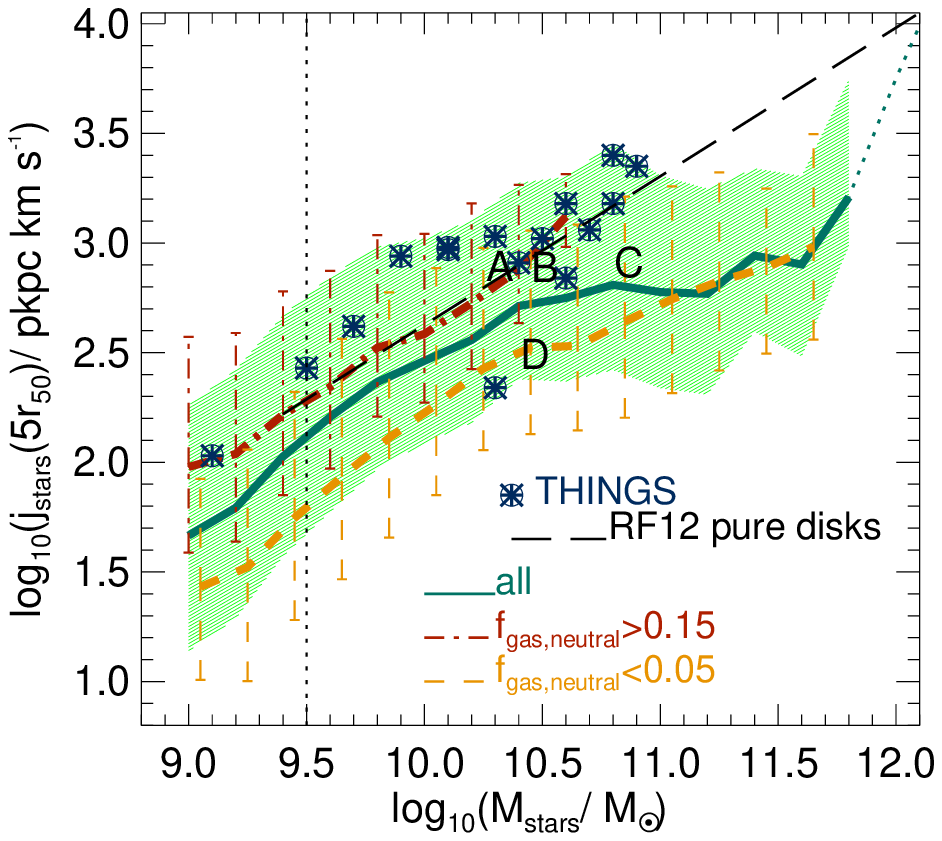}
\includegraphics[width=0.355\textwidth,valign=t]{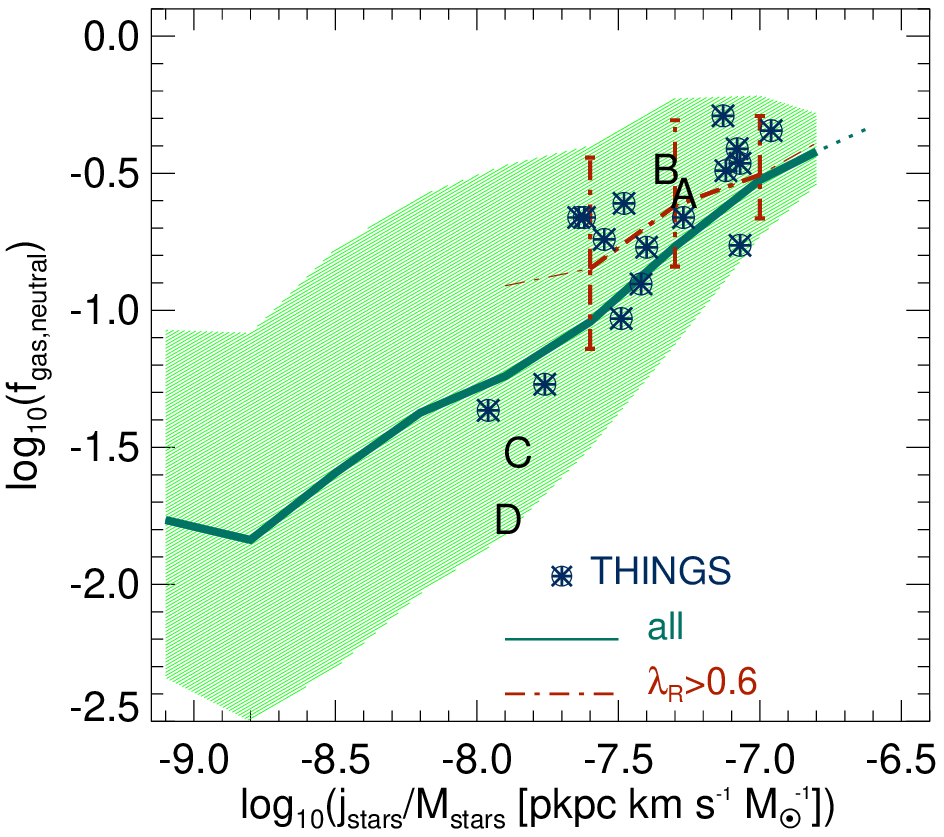}
\includegraphics[width=0.28\textwidth,valign=t]{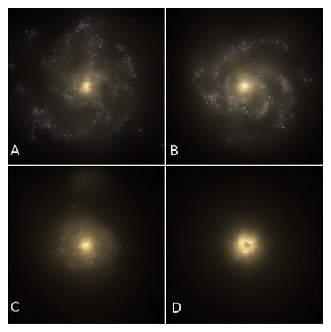}
\caption{{\it Left panel:} $j_{\rm stars}$, measured within $5$ times the half-mass radius of the stellar component, as a function of stellar mass, 
for galaxies at $z=0$ with $M_{\rm stars}>10^9\,\rm M_{\odot}$ (no restriction in $\lambda_{\rm R}$ is applied here). 
The solid line and shaded region show the
median and $16^{\rm th}$ to $84^{\rm th}$ percentile range, respectively.
We also show the median and $16^{\rm th}$ to $84^{\rm th}$ percentile range of the relation for the subsample of galaxies 
with $f_{\rm gas,neutral}>0.15$ and $f_{\rm gas,neutral}<0.05$, as lines with error bars. 
Bins that with $<10$ objects are shown as dotted lines.
Observational results from \citet{Obreschkow14b} using the THINGS survey are shown as star symbols.
We also show the observational result of \citet{Romanowsky12} for disks.
For reference, the vertical line shows a conservative stellar mass limit above which $j_{\rm stars}$ is well converged for the resolution of the simulation. 
{\it Middle panel:} As in the left panel, but here we show the neutral gas fraction, $f_{\rm gas,neutral}$, as a function of $j_{\rm stars}/M_{\rm stars}$.
The relation for the subsample of galaxies with $\lambda_{\rm R}>0.6$ is shown as the dot-dashed line with error bars
(median and $16^{\rm th}$ to $84^{\rm th}$ percentile range, respectively).
{\it Right panel:} SDSS gri face-on images of $4$ \eagle\ galaxies with 
$M_{\rm stars}=10^{10.3}-10.0^{10.7}\,\rm M_{\odot}$ and
$f_{\rm gas,neutral}>0.15$ (top images) or $f_{\rm gas,neutral}<0.05$ (bottom images), constructed using the radiative transfer
code {\tt SKIRT} \citep{Baes11} (Trayford et al. in preparation). The images are $60\,\rm pkpc$ on a side 
and the positions of these galaxies, A, B, C and D, in the left and middle panels 
are shown with the corresponding. These images are publicly available from the 
\eagle\ database \citep{McAlpine15}. The figure shows that gas-rich galaxies have significantly higher $j_{\rm stars}$ 
and agree better with the THINGS observations. These galaxies appear visually to be similar to the 
late-type galaxies observed by THINGS.} 
\label{JMcomp2}
\end{center}
\end{figure*}

Here we compare the predictions of \eagle\ with four sets of observations: 
the \citet{Romanowsky12} sample, the ATLAS$^{\rm 3D}$ survey \citep{Cappellari11}, the SAMI survey \citep{Croom12} and the THINGS
survey \citep{Walter08,Obreschkow14b}. Below we give a brief overview of how $j_{\rm stars}$ was calculated in the four datasets used here.

\begin{itemize}
\item {\it \citet{Romanowsky12}.} This corresponds to a sample of $\approx 100$ galaxies. Unlike the other observational samples we use here, 
measurements from \citet{Romanowsky12} were not done using resolved kinematic information, but instead they use long-slit spectroscopy and 
 the HI emission line. This means that these measurements are considered to be {\it total} stellar specific angular momentum. 

\item {\it ATLAS$^{\rm 3D}$.} In order to calculate the stellar angular momentum within the
effective (half-light) radius of the ATLAS$^{\rm 3D}$ early-type galaxies (ETGs), we retrieved the stellar
kinematics for all $260$ objects derived in \citet{Cappellari11}. Following \citet{Obreschkow14b}, we correct the projected velocity observed in
each spaxel of the IFU for inclination by assuming a thin disc model for
each object, with the position angle derived in \citet{Krajnovic11} and the inclination from \citet{Cappellari13}. 
Spaxels very close to the minor axis of the galaxy were
blanked, to avoid numerical artefacts. From these de-projected velocities
we calculate $j_{\rm stars}$ within the effective radius (taken from \citealt{Cappellari11}) following the equations in
\citet{Obreschkow14b}.
We note that a thin disk model may not be appropriate for ETGs, which
can have significant bulge components. As ATLAS$^{\rm 3D}$ have shown that $\approx 86$\% of
these objects are fast rotators \citep{Emsellem11}, with embedded stellar discs \citep{Krajnovic12,Krajnovic13}
and {axisymmetric rotation curves \citep{Cappellari11b}}, 
we do not expect this procedure to yield significant bias. In addition, 
{\citet{Naab14} showed that fast rotators in simulations have velocity moments that are consistent with             
disks.}
However, results for slow rotators
should be treated carefully, {as this approximation is likely to be inappropriate in that regime.}
Measurements in ATLAS$^{\rm 3D}$ were done in circularised 
effective radii. The latter is $\approx 1.4$ times smaller than for example the ones used in SAMI (described below) 
at fixed stellar mass (and for the same morphological type). Thus, to compare \eagle\ with ATLAS$^{\rm 3D}$ we therefore need 
to produce a similar estimate of a circularised, 2-dimensional projected $r_{50}$ and then measure $j_{\rm stars}$ within that aperture.
We call the latter radius $r_{\rm 50,circ}$.

\item {\it SAMI.} \citet{Cortese16} presented the 
measurements of the $j_{\rm stars}-M_{\rm stars}$ relation for galaxies of different morphological types and different values of 
$\lambda_{\rm R}$, in the stellar mass range $10^{9.5}\,\rm M_{\odot}\lesssim M_{\rm stars}\lesssim 10^{11.4}\,\rm M_{\odot}$. 
Cortese et al. measured $j_{\rm stars}$ within an effective radius from the 
line-of-sight velocity measured in each spaxel, and following the optical ellipticity and position 
angle of galaxies. These measurements are then corrected for inclination. Note that 
here the effective radius is similar to how we measure $r_{50}$ in \eagle\, and thus we can directly compare 
the results presented in $\S$~\ref{jz0sec} with SAMI.

\item {\it THINGS.} In the case of the THINGS survey, 
\citet{Obreschkow14b} presented a measurement of the $j_{\rm stars}$-stellar mass relation, where $j_{\rm stars}$ was measured within 
$\approx 10$ times the scale radius, which for an exponential disk, corresponds to $\approx 5$ times the half-mass radius. These represent the most 
accurate measurements of $j_{\rm stars}$ and $j_{\rm bar}$ to date, owing to the very high resolution and depth of the dataset used by 
\citet{Obreschkow14b}. These measurements are not comparable to those of ATLAS$^{\rm 3D}$ and SAMI, given that the latter only probe
$j$ within $r_{\rm 50}$. 
\end{itemize}

In Fig.~\ref{JMcomp} we present 
the $j_{\rm stars}(r_{50})$-stellar mass relation in three bins of $\lambda_{\rm R}$, $<0.3$, $0.3-0.5$ and $>0.5$.
The predicted relations here are much tighter than the one shown in the top panel of Fig.~\ref{JMEAGLE}, with a scatter of 
$\approx 0.15-0.3$~dex, which is a consequence of the limited range of $\lambda_{\rm R}$ studied in each panel.

The top panel of Fig.~\ref{JMEAGLE} shows galaxies with $\lambda_{\rm R}<0.3$ in both simulation and observations. 
{SAMI galaxies are shown as symbols.}
\eagle\ is in broad agreement with SAMI, although with a slightly smaller median than that of SAMI galaxies.
%{The latter could be partially due to $r_{\rm 50}$ in \eagle\ being intrinsic, while being projected in SAMI.}
The predicted $1\sigma$ dispersion in \eagle\ is also similar to the one measured in SAMI, which is $\approx 0.26$~dex \citep{Cortese16}. 
Here we also show the approximate location of the observational results of \citet{Romanowsky12}
for bulges and found that they are on the upper envelope of both SAMI and \eagle.
This is not surprising given that \citet{Romanowsky12} presented measurements of the total $j_{\rm stars}$.
In the middle panel of Fig.~\ref{JMcomp} we show galaxies with $0.3\le \lambda_{\rm R}<0.5$. 
The observations of SAMI show that the increase in normalisation once higher $\lambda_{\rm R}$ galaxies 
are selected is very similar to the increase obtained in \eagle. 
The bottom panel of Fig.~\ref{JMcomp} shows galaxies with $\lambda_{\rm R}\ge 0.5$. 
Here, SAMI galaxies lie slightly above the \eagle\ galaxies at 
$M_{\rm stellar}\gtrsim 10^{10.7}\rm\,M_{\odot}$, although well within the 
$1\sigma$ dispersion in both samples. 
Also shown is the approximate location of the observational results of \citet{Romanowsky12} 
for disks. {The slope of this sample is slightly steeper than what we obtain for \eagle\ galaxies.}
The results of Fig.~\ref{JMcomp} are consistent with  
\eagle\ and SAMI galaxies {spanning a continuous sequence in the $j_{\rm stars}$-stellar mass plane, that go from low $j$-low $\lambda_{\rm R}$-low $V_{\rm rot}/\sigma_{\rm stars}$ to 
high $j$-high $\lambda_{\rm R}$-high $V_{\rm rot}/\sigma_{\rm stars}$}. 

To compare with ATLAS$^{\rm 3D}$ we measure $j_{\rm stars}$ inside the circularised, 2-dimensional half-mass radius of the stellar component, 
$r_{\rm 50,circ}$. We do this by taking the projected stellar mass map on the $x-y$ plane and measuring the half-mass radius in circular 
apertures. In addition, we select galaxies in \eagle\ that are passive, 
which would match well the properties of ATLAS$^{\rm 3D}$ ETGs (mostly passive, except for a couple of galaxies). 
We use two selections: (1) \eagle\ galaxies with a specific SFR (sSFR) $<0.01\,\rm Gyr^{-1}$, which would select 
galaxies below the main sequence in the $\rm SFR-M_{\rm stars}$ plane (\citealt{Furlong15}), and (2) 
\eagle\ galaxies with $\rm (u^*-r^*)>2$, which selects galaxies in the red sequence (\citealt{Trayford16}).
We compare the above subsamples with ATLAS$^{\rm 3D}$ in Fig.~\ref{JMcompA3D}. 
We find that both subsamples of \eagle\ galaxies agree very well with 
the measurements, albeit with \eagle\ possibly predicting a slightly shallower relation. However, the difference is well within the uncertainties. 
The scatter of \eagle\ is slightly larger than that found in ATLAS$^{\rm 3D}$ ($0.6$ vs. $0.4$~dex, respectively). 
This may be due to the lack of a true morphological selection of galaxies in \eagle\, which would require a visual 
inspection of the synthetic $gri$ images.
In addition, there are some ATLAS$^{\rm 3D}$ galaxies with very low $j_{\rm stars}$ measurements. These are the slow rotators, 
and it is likely that our estimates are systematically lower in these objects because our disc assumption is not valid in this regime.

In the left panel of Fig.~\ref{JMcomp2} we show the $j_{\rm stars}-M_{\rm stars}$ relation with $j_{\rm stars}$ now measured within $5\,r_{5}$, 
for galaxies with $M_{\rm stars}>10^9\,\rm M_{\odot}$ at $z=0$ in \eagle.
Individual measurements from \citet{Obreschkow14b} are shown as symbols. Here we show again the approximate 
location of the observational results of \citet{Romanowsky12} 
for disks. The observations of \citet{Obreschkow14b} are well within the scatter of the relation of all \eagle\ galaxies, but the 
median of the simulation is systematically offset by $\approx 0.2$~dex to lower values of $j_{\rm stars}$. 
At $M_{\rm stars} \gtrsim 10^{10.3}\rm\, M_{\odot}$, \eagle\ galaxies systematically deviate from the observations of 
 \citet{Obreschkow14b} and \citet{Romanowsky12}. 
To reveal the cause of the offset, we 
divide the \eagle\ sample into gas-rich ($f_{\rm gas,neutral}>0.15$) and gas-poor ($f_{\rm gas,neutral}<0.05$) galaxies, 
and present the median and scatter of those sample as dot-dashed and dashed lines with error bars, respectively. 
The subsample of galaxies with $f_{\rm gas,neutral}>0.15$ shows no flattening of the $j_{\rm stars}-M_{\rm stars}$ relation 
and the median is shifted upwards to higher $j_{\rm stars}$.
 The sample of \citet{Obreschkow14b} is characterised by a 
median $f_{\rm gas,neutral}\approx 0.22$, meaning that it should be compared to 
the \eagle\ sample with $f_{\rm gas,neutral}>0.15$. By doing this, we find excellent agreement between 
\eagle\ and the THINGS observations. {Thus, the $j_{\rm stars}-M_{\rm stars}$ relation found by 
\citet{Obreschkow14b} is not representative of the overall galaxy population at fixed stellar mass, but only 
of the relatively gas-rich galaxies.}

To help visualise how gas-rich vs. gas-poor galaxies of the same stellar mass look like, we show 
SDSS gri face-on images of four \eagle\ galaxies with stellar masses in the range $10^{10.3}-10^{10.7}\,\rm M_{\odot}$, and 
$f_{\rm gas,neutral}>0.15$ (galaxies A and B, top images) or 
$f_{\rm gas,neutral}<0.05$ (galaxies C and D, bottom images). These images were created using radiative transfer simulations
performed with the code {\tt SKIRT} \citep{Baes11} in the SDSS g,~r and i filters \citep{Doi10}. Dust extinction was implemented
using the metal distribution of galaxies in the simulation, and assuming $40$\% of the metal mass is locked up in dust grains \citep{Dwek98}. 
The images were produced
using particles in spherical apertures of $30$~pkpc around the centres of sub-halos (see \citealt{Trayford15}, and in preparation for more details).
It is clear that galaxies that look like regular spiral galaxies in \eagle\ correspond to those having 
$f_{\rm gas,neutral}\gtrsim 0.15$, while gas-poor galaxies look like lenticulars or early-type galaxies. 
Galaxies C and D have $\lambda_{\rm R}(r_{\rm 50})\approx 0.45$, which  
 would be classified observationally as fast rotators early-type galaxies in the nomenclature 
of the ATLAS$^{\rm 3D}$ survey \citep{Cappellari11}. The positions of these galaxies in the 
$j_{\rm stars}$-$M_{\rm stars}$ plane are shown in the left and middle panel of Fig.~\ref{JMcomp2} with the corresponding letters.
We visually inspected $100$ galaxies randomly selected 
($50$ in the gas-rich and $50$ in the gas poor subsamples), and found the differences presented 
here (between galaxies A-B and C-D) to be generic.

We further characterise the relation between $j_{\rm stars}$, stellar mass and $f_{\rm gas,neutral}$ 
{in the middle panel of Fig.~\ref{JMcomp2}}, that shows $f_{\rm gas,neutral}$ vs. $j_{\rm stars}/M_{\rm stars}$ 
for \eagle\ galaxies at $z=0$ with 
$M_{\rm stars}>10^{9.5}\,\rm M_{\odot}$. In \eagle\ galaxies that are more gas rich, also have a higher 
$j_{\rm stars}/M_{\rm stars}$. The scatter is slightly reduced if we select galaxies in narrow ranges of 
$\lambda_{\rm R}$ (see dot-dashed line with error bars in the middle panel of Fig.~\ref{JMcomp2}). The observations of \citet{Obreschkow14b} 
fall within the $1\sigma$ scatter of the relation in \eagle, which shows that the simulation captures how the angular momentum together with 
the gas content of galaxies are acquired.

The agreement between the simulation and the observations is quite remarkable. 
\eagle\ not only reproduces the normalisation of the $j_{\rm stars}$-stellar mass relation, 
which may not be so surprising given that \eagle\ matches the size-stellar mass relation well \citep{Schaye14,Furlong15}, 
but also the trends with $\lambda_{\rm R}$ and $f_{\rm gas,neutral}$ as identified by observations.

\section{The evolution of the specific angular momentum of galaxies in \eagle}\label{StructureEvolution}

Here we analyse the evolution of $j_{\rm stars}$ and $j_{\rm neutral}$ as a function of galaxy properties and 
 attempt to find those properties that are more fundamentally correlated to them. 

\subsection{The evolution of the $j$-mass relations}\label{evojsec}

\begin{figure}
\begin{center}
\includegraphics[width=0.45\textwidth]{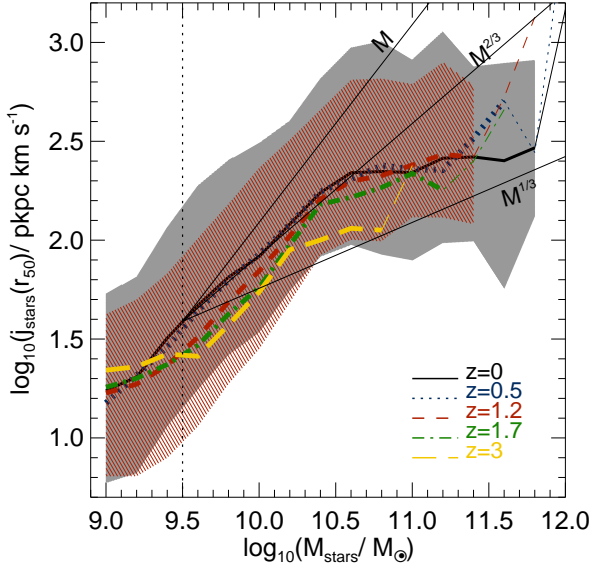}
\caption{The specific angular momentum of the stars measured with all the particles within the half-mass radius of the stellar component as a function
of stellar mass at $z=0$, $0.5$, $1.2$, $1.7$ and $3$, as labelled, for all galaxies with $M_{\rm stars}>10^9\,\rm M_{\odot}$ and 
$r_{rm 50}>1$~pkpc in \eagle. Lines show the median relations, while the shaded regions show the 
$16^{\rm th}$ to $84^{\rm th}$ percentile ranges, and are only shown for $z=0,\,1.2$. 
For reference, the vertical dotted line shows a conservative stellar mass limit above which $j_{\rm stars}$ 
is well converged for the resolution of the simulation,
and the straight solid lines show the scalings $j\propto M$, $j\propto M^{2/3}$ and $j\propto M^{1/3}$, as labelled.}
\label{JVSMsEvoa}
\end{center}
\end{figure}

\begin{figure}
\begin{center}
\includegraphics[width=0.45\textwidth]{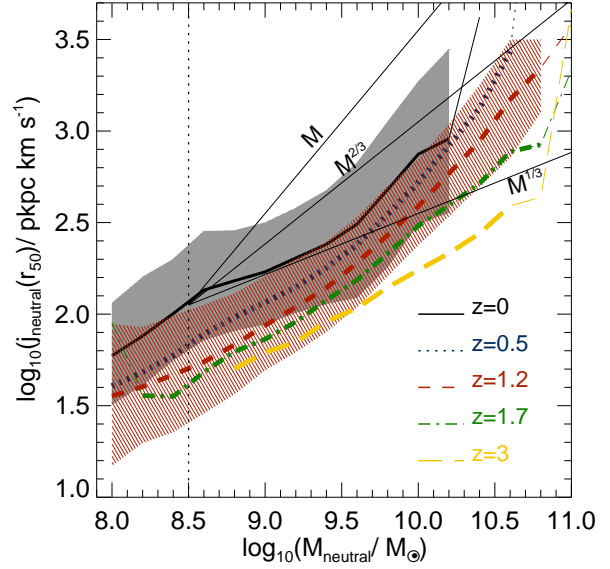}
\caption{As in Fig.~\ref{JVSMsEvoa}, but here we show $j_{\rm neutral}$ as a function of
neutral gas mass. At fixed mass, galaxies at high redshift have lower $j_{\rm neutral}$ than the $z=0$
counterparts.}
\label{JVSMsEvoc}
\end{center}
\end{figure}

\begin{figure}
\begin{center}
\includegraphics[width=0.45\textwidth]{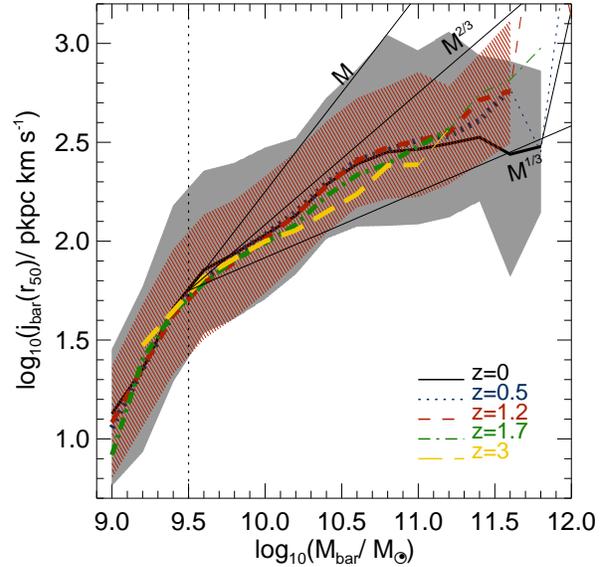}
\caption{As in Fig.~\ref{JVSMsEvoa}, but here we show $j_{\rm bar}$ as a function of 
baryon mass (stars plus neutral gas).}
\label{JVSMsEvob}
\end{center}
\end{figure}

\begin{figure*}
\begin{center}
\includegraphics[width=0.77\textwidth]{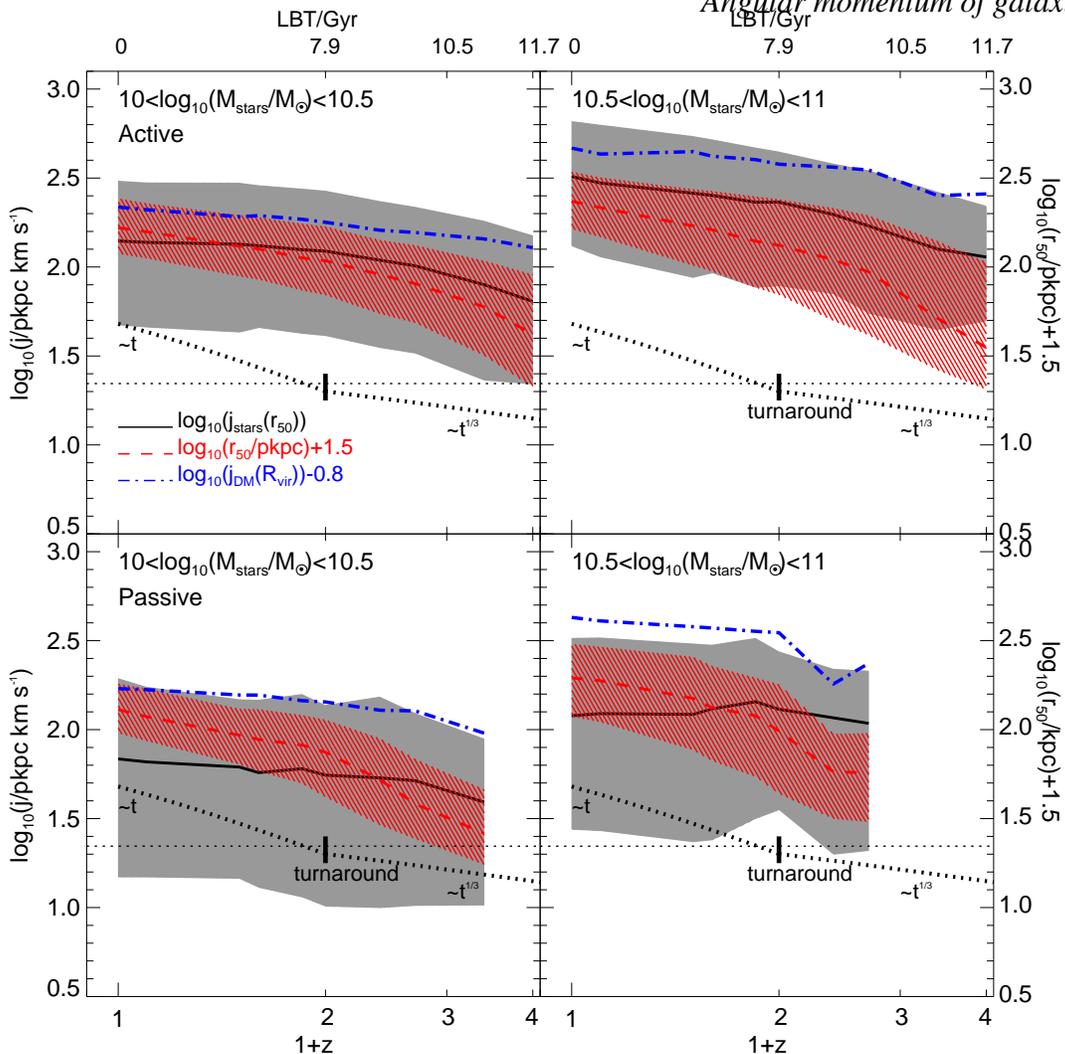}
\caption{Evolution of $j_{\rm stars}(r_{50})$ 
for {central} galaxies in \eagle\ in two bins of stellar mass, as labelled in each panel, and separated into 
active (top panels) and passive (bottom panels) galaxies. The latter classification is made 
based on their position with respect to the main sequence.
{Look-back time is shown at the top.}
The line and shaded region shows the median and $16^{\rm th}-84^{\rm th}$ percentile range 
of $j_{\rm stars}(r_{50})$, respectively. Only bins with $\ge 10$ galaxies are shown. 
The $j$ of the host DM halo, {$j_{\rm DM}(\rm R_{\rm vir})$ (scaled by $-0.8$~dex) 
is shown as dot-dashed lines.} 
We also show the evolution of half-mass radius of the stellar component, $r_{50}$, plus $1.5$~dex to match the normalisation 
of $j_{\rm stars}$ (median and $16^{\rm th}-84^{\rm th}$ percentile range shown as dashed line with shaded region, 
respectively).
In each panel we show for reference the maximum gravitational softening length of the simulation (plus $1.5$~dex) 
as horizontal dotted line (at $z\ge 2.7$ it decreases as $(1+z)$, but for clarity we do not show that here). 
The vertical segments show roughly the turnaround epoch of the host halos.
In addition, we show as dotted line (using an arbitrary normalisation) the prediction of \citet{Catelan96a} 
of how $j_{\rm stars}$ grows with time before and after the turnaround epoch. {The latter is only for reference, and should not be 
taken as a test of the theoretical predictions, given that here we are not tracing galaxy progenitors, but instead 
selecting similar galaxy populations at different redshifts.} 
This figure shows that galaxy sizes evolve much more strongly than $j_{\rm stars}$ at fixed stellar mass, 
{and that star-forming galaxies exhibit a stronger increase in $j_{\rm stars}$ than passive galaxies}.}
\label{JVSMsR50Evo}
\end{center}
\end{figure*}

\begin{figure}
\begin{center}
\includegraphics[width=0.45\textwidth]{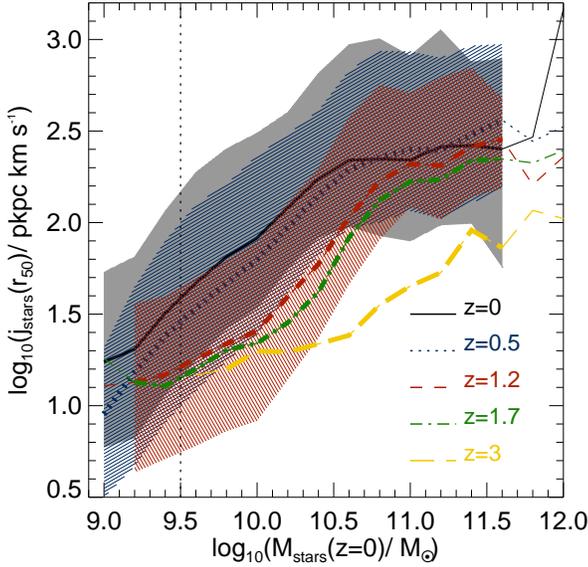}
\caption{The specific angular momentum of the stars measured at different redshifts as a function of the stellar 
mass galaxies have at $z=0$. In the case of $j$, we measure it at $z=0$, $0.5$, $1.2$, $1.7$ and $3$, as labelled, 
with all the particles within the half-mass radius of the stellar component at the redshift. We show the relation 
for all galaxies with $M_{\rm stars}>10^9\,\rm M_{\odot}$ at $z=0$ 
in \eagle. Lines show the median relations, while the shaded regions show the
$16^{\rm th}$ to $84^{\rm th}$ percentile ranges, and are only shown for $z=0,\,0.5,\,1.2$. 
Bins with $<10$ objects are shown as thin lines. This figure shows that 
progenitor galaxies typically have lower $j$ than their descendants. 
For reference, the vertical line shows a conservative stellar mass limit above which $j_{\rm stars}$ is well converged for the resolution of the simulation.}
\label{SigmaEvoAscendant}
\end{center}
\end{figure}

\begin{figure}
\begin{center}
\includegraphics[width=0.45\textwidth]{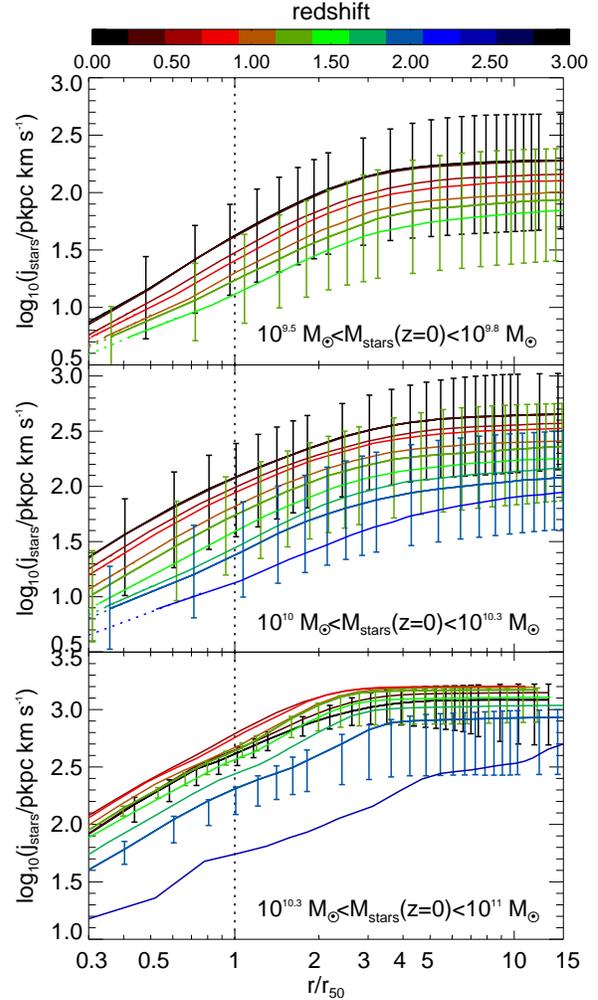}
\caption{{$j_{\rm stars}$ measured within $r$ as a function of $r$ in units of $r_{\rm 50}$ at different 
redshifts, as shown by the color bar, for galaxies with $z=0$ stellar mass in three bins, as labelled in each panel.
Solid lines show the median, while the error bars show the $16^{\rm th}$ to $84^{\rm th}$ percentile ranges. 
The latter are only shown for $z=0,\,1,\,1.7$. Dotted lines show the extrapolation of the profiles towards radii that are below 
$1$~pkpc (approximately the spatial resolution of \eagle). The vertical dotted lines shows $r=r_{\rm 50}$. The range span by the y-axis changes in each panel 
to better cover the dynamic range of $j_{\rm stars}$ in each stellar mass bin.
This figure shows that $j_{\rm stars}(r_{\rm 50})$ evolves more strongly than $j_{\rm stars}$ measured 
at larger apertures.}}
\label{RadProfilesJstars}
\end{center}
\end{figure}

\begin{figure}
\begin{center}
\includegraphics[width=0.48\textwidth]{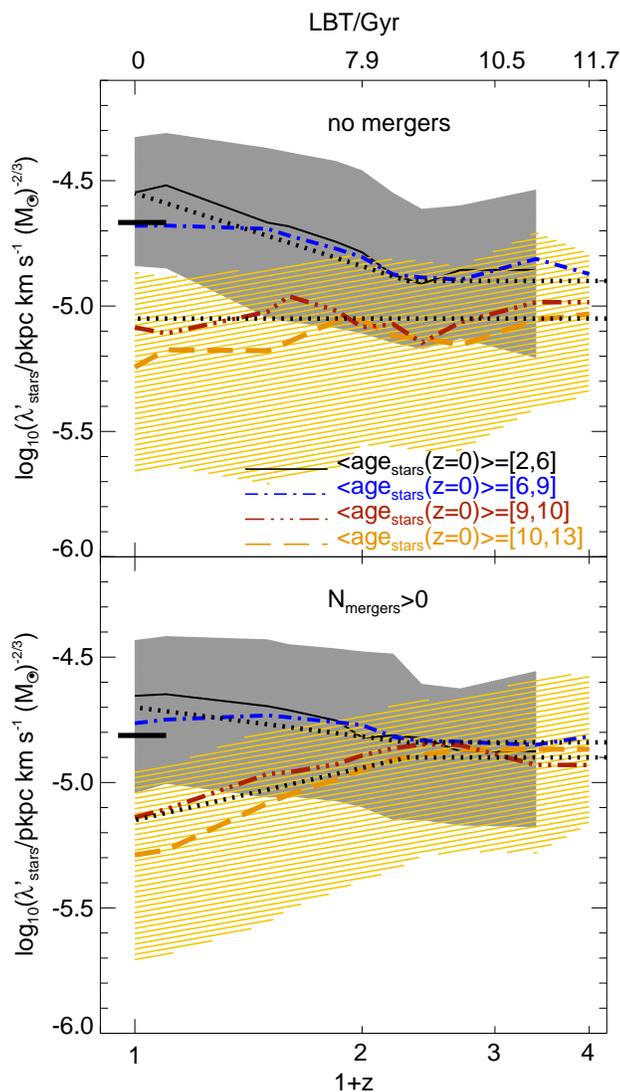}
\caption{The value of $\lambda^{\prime}_{\rm stars}\equiv j_{\rm stars}(r_{50})/M^{2/3}_{\rm stars}$ as a function of redshift, for \eagle\ galaxies 
selected in different bins of mass-weighted stellar age at $z=0$, $\langle\rm age_{\rm stars}\rangle$, 
as labelled (with numbers in the figure being in Gyr), and that have 
$M_{\rm stars}(z=0)\ge 10^{9.5}\,\rm M_{\odot}$. Look-back time is shown at the top. Lines show the median relations, and 
the shaded regions show the $25^{\rm th}$ to $75^{\rm th}$ percentile ranges, and for clarity we only show 
this for the lowest and highest $\rm \langle\rm age_{\rm stars}\rangle$ bins. In the top panel we show the subsamples of galaxies 
at $z=0$ that had not suffered galaxy mergers, while the bottom panel shows the complement. 
The segment in both panels show the median of the selected galaxy population at $z=0$, while the dotted 
lines show the average evolutionary tracks {of Table~\ref{EvoTracksAve}.} }
\label{SigmaEvoAscendantAge}
\end{center}
\end{figure}

\begin{figure}
\begin{center}
\includegraphics[width=0.45\textwidth]{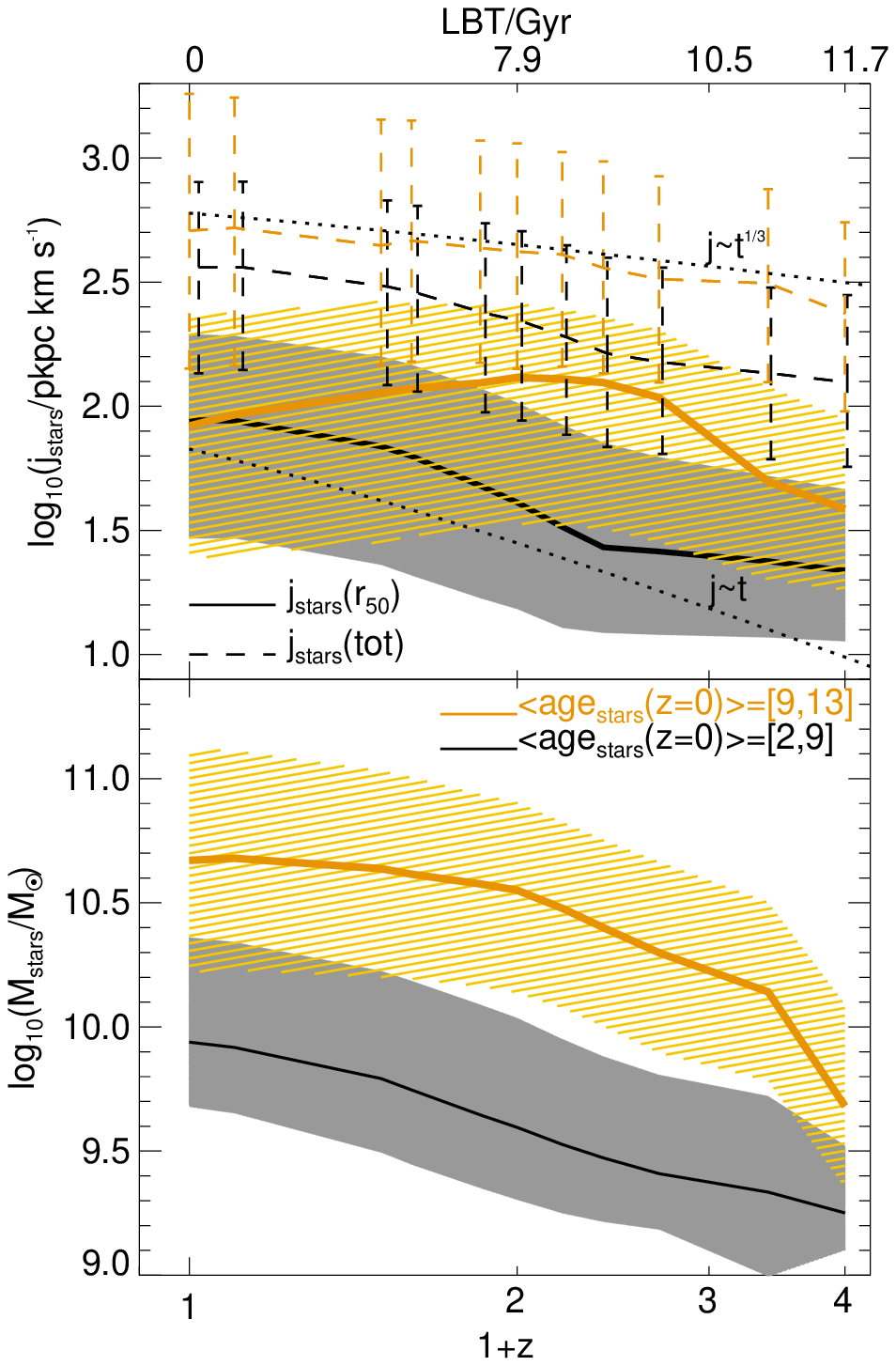}
\caption{{{\it Top panel:} Evolution of the median $j_{\rm stars}(r_{50})$ (solid lines) and 
$j_{\rm stars}(\rm tot)$ (dashed lines) for two samples of 
galaxies selected by their mass-weighted stellar age at $z=0$, 
as labelled in the bottom panel. 
$j_{\rm stars}(\rm tot)$ is measured with all the star particles 
in the sub-halo, while $j_{\rm stars}(r_{50})$ only with those within $r_{\rm 50}$. 
Lines show the medians, while the shaded regions and error bars show the 
$25^{\rm th}$ to $75^{\rm th}$ percentile ranges.
The dotted lines (using an arbitrary normalisation) show the prediction 
{of \citet{Catelan96a,Catelan96b} of how $j$ grows with time before the turnaround epoch ($j\propto t^{1/3}$), and 
the upper limit for the time dependence of the specific angular momentum of infalling material after turnaround ($j\propto t$). 
The latter could therefore be considered as an upper limit for how fast $j_{\rm stars}(\rm tot)$ can increase.}
{\it Bottom panel:} 
Evolution of the median stellar mass of the galaxies shown in the top panel. 
This figure shows that galaxies throughout their lifetimes go through a significant rearrangement 
of their $j_{\rm stars}$ in a way that young galaxies have inner $j_{\rm stars}$ growing faster than the total value, while 
old galaxies at $z\lesssim 1$, have inner $j_{\rm stars}$ decreasing, while their total $j_{\rm stars}$ shows little evolution.}}
\label{SigmaEvoAscendantAgeProperties}
\end{center}
\end{figure}

Fig.~\ref{JVSMsEvoa} shows the $j_{\rm stars}(r_{\rm 50})-M_{\rm stars}$ relation in the redshift range 
$0\le z\le 3$ for galaxies with $M_{\rm stars}>10^9\,\rm M_{\odot}$ and $r_{\rm 50}>1$~pkpc in \eagle. Galaxies have lower 
$j_{\rm stars}(r_{\rm 50})$ at fixed stellar mass at high redshift. Interestingly, between $0\lesssim z\lesssim 0.5$ the 
normalisation of the $j_{\rm stars}(r_{\rm 50})-M_{\rm stars}$ relation evolves weakly. The strongest change experienced by galaxies is at 
$1\le z \le 3$ (of $\approx 0.2-0.35$~dex).
 The stellar mass above which the $j_{\rm stars}(r_{\rm 50})-M_{\rm stars}$ relation flattens has 
a small tendency of increasing with decreasing redshift. 
At $z\approx 1.2$ the flattening is seen above $\approx 10^{10.3}\,\rm M_{\odot}$, 
while at $z=0$ the flattening starts at $10^{10.5}\,\rm M_{\odot}$.
There are no available observational measurements of $j_{\rm stars}(r_{\rm 50})$ at high redshift yet, but 
there are measurements of how the effective radius and the rotational velocity of galaxies evolve. 
\citet{vanderwel14} showed that galaxies at fixed stellar mass are $\approx 1.9$ times smaller 
at $z=1$ compared to $z=0$, while in the same redshift range \citet{Tiley16} showed that galaxies 
increase their rotational velocity by $\approx 1.3$. If one assumes 
that $j_{\rm stars} \sim r_{50}\,V_{\rm rot}$, then these observations imply 
a decrease of $j_{\rm stars}$ at fixed stellar mass from $z=0$ to $z=1$ of $\approx 1.4-1.5$, very similar to the 
magnitude of evolution in $j_{\rm stars}$ we obtain from \eagle\ at $0\le z\le 1.2$.
 
The flattening of the $j_{\rm stars}(r_{\rm 50})-M_{\rm stars}$ relation at high stellar 
masses is mostly driven by galaxy mergers (Fig.~\ref{JsJbEvoMergers}).
In Fig.~\ref{JVSMsEvoa} we also show for comparison the scalings 
$j\propto M$, $j\propto M^{2/3}$ and $j\propto M^{1/3}$. 
A scaling $j\propto M$ is expected in the model of \citet{Obreschkow14b}, where galaxies 
are well described by the relation $Q\propto j_{\rm stars}\,M^{-1}_{\rm stars}\,(1-f_{\rm gas,neutral})\,\sigma$, 
while a relation $j\propto M^{2/3}$ 
is predicted in a CDM universe under the assumption of conservation of $j$ ($\S$~\ref{theoryback}).
Galaxies with stellar masses below the flattening 
{and at $z\lesssim 1$ follow a scaling close to $j_{\rm stars}\propto M^{2/3}_{\rm stars}$, while 
at higher redshifts the relation becomes steeper, which is most evident in the mass range 
$10^{9.4}\,\rm M_{\odot}\lesssim M_{\rm stars}\lesssim 10^{10.5}\,\rm M_{\odot}$.}
{By fitting the $j_{\rm stars}(r_{\rm 50})-M_{\rm stars}$ relation using a power-law and 
the {\tt HYPER-FIT} R package of \citet{Robotham15} we find 
 that the best fit power-law index at $z\gtrsim 1$ in the stellar mass range above is $\approx 0.77$.}

Fig.~\ref{JVSMsEvoc} shows the $j_{\rm neutral}(r_{\rm 50})-M_{\rm neutral}$
relation for galaxies with $M_{\rm stars}>10^9\,\rm M_{\odot}$ and $r_{\rm 50}>1$~pkpc at $0\le z\le 3$ in \eagle.
Galaxies evolve significantly in this plane, having $\approx 3-5$ times lower $j_{\rm neutral}$
at $z\approx 3$ than they do at $z=0$ at fixed
$M_{\rm neutral}$.  
{By fitting the $j_{\rm neutral}-M_{\rm neutral}$ relation using {\tt HYPER-FIT} 
we find that the best fit power-law index is $\approx 0.5-0.6$, with the exact value depending on the redshift. 
Thus, on average, this relation is close to the theoretical expectation of $j\propto M^{2/3}$.}

In Fig.~\ref{JVSMsEvob} 
we show the $j_{\rm bar}(r_{\rm 50})-M_{\rm bar}$ relation for galaxies 
with $M_{\rm bar}>10^9\,\rm M_{\odot}$ and $r_{\rm 50}>1$~pkpc at $0\le z\le 3$ in \eagle.
{There is little evolution of the $j_{\rm bar}-M_{\rm bar}$ at $M_{\rm bar}\lesssim 10^{10}\,\rm M_{\odot}$. 
Galaxies with $10^{10}\,\rm M_{\odot}\lesssim M_{\rm bar}\lesssim 10^{11}\,\rm M_{\odot}$, 
display a modest evolution of $j_{\rm bar}$ at $1\lesssim z\lesssim 3$ of $\approx 0.2$~dex, with little 
evolution below $z\sim 1$.
Galaxies with $M_{\rm bar}\gtrsim 10^{11}\,\rm M_{\odot}$ have $j_{\rm bar}(r_{\rm 50})$ decreasing at fixed stellar mass. The latter             
is due to galaxies becoming increasingly gas poor, and thus going from $j_{\rm bar}$ being dominated by $j_{\rm neutral}$ to being 
dominated by $j_{\rm stars}$.
The former is almost always higher than the latter.}
Note that the power-law index of the $j_{\rm bar}$-baryon mass does not change significantly with redshift and is always close to 
$\approx 0.6$, {although noticeable differences are seen with stellar mass, at fixed redshift.} 

\subsubsection{$j_{\rm stars}$ evolution in active and passive galaxies}

{Fig.~\ref{JVSMsR50Evo} shows the evolution of $j_{\rm stars}(r_{\rm 50})$ for active and passive central galaxies in two 
bins of stellar mass. We select central galaxies to enable us to compare with $j_{\rm DM}(\rm R_{\rm vir})$, 
which is calculated with {\it all} DM particles within the virial radius of the friends-of-friends host halo. 
For comparison, we also show the evolution of $r_{50}$.} 
We separate galaxies into active and passive by calculating the position of the main sequence at each redshift, and then computing 
the distance in terms of sSFR
to the main sequence, $\rm sSFR/\langle sSFR_{\rm MS}\rangle$. 
Galaxies with $\rm sSFR/\langle sSFR_{\rm MS}\rangle<0.1$ are considered passive, while the complement are active. 
The position of the main sequence, $\rm \langle sSFR_{\rm MS}\rangle$, is calculated as the median sSFR of all galaxies at a given redshift that 
have $\rm sSFR>sSFR_{\rm lim}$, where $\rm sSFR_{\rm lim}$ is defined as 
$\rm log_{10}(sSFR_{\rm lim}/Gyr^{-1})=0.5\,z-2$ for $z\le 2$ and $\rm log_{10}(sSFR_{\rm lim}/Gyr^{-1})=-1$ 
for $z>2$ (see \citealt{Furlong14} for details).
 
Once the stellar mass is fixed, 
$j_{\rm stars}(r_{\rm 50})$ evolves very weakly in passive galaxies ($\approx 0.2$~dex between $0\le z\le 3$) 
and slightly more strongly in star-forming galaxies ($\approx 0.3-0.4$~dex between $0\le z\le 3$).
{We show the evolution of $j_{\rm DM}(R_{\rm vir})$ of the halo hosting the galaxies shown in Fig.~\ref{JVSMsR50Evo}, 
to stress the fact that the evolution of $j_{\rm stars}\sim j_{\rm DM}$, to within $\approx 50$\% (i.e. 
the $1\sigma$ scatter around the constant of proportionality is $\approx 0.18$~dex).
We remind the reader that we are not studying the evolution of individual galaxies here, but instead how $j$ evolves at fixed stellar mass and 
star formation activity, as defined by the distance of galaxies to the main sequence of star formation.} 
The selection of galaxies in Fig.~\ref{JVSMsR50Evo} roughly corresponds to halos of the same mass at different redshifts. { 
At fixed mass, halos also show a slight increase in $j_{\rm DM}$ with time due to halos at lower redshift crossing turnaround 
at increasingly later times, which imply that they had longer times to acquire angular momentum}. 
{The similarity seen between $j_{\rm stars}(r_{50})$ and $j_{\rm DM}(\rm R_{\rm vir})$} 
means that to {zeroth} order any gastrophysics is 
secondary when it comes to the value of $j_{\rm stars}$ in galaxies, showing how fundamental this quantity is.
{However, when studied in detail, we find that galaxies undergo a significant rearrangement of their $j_{\rm stars}$ radial profile that 
is a result of galaxy formation. This rearrangement is also the cause of $j_{\rm DM}$ evolving much more weakly 
than $j_{\rm stars}(r_{\rm 50})$, particularly in star-forming galaxies. We come back to this in $\S$~\ref{TraceBackj}.}

The evolution of $j_{\rm stars}(r_{\rm 50})$ {at fixed stellar mass} in \eagle\ mostly occurs at $z\gtrsim 1$, 
{before the turnaround epoch of the halos hosting the
galaxies of the stellar mass we are studying here, at $z=0$, which is 
$z \approx 1$.}
This epoch corresponds to the time of {\it maximum expansion} that is followed by the collapse of halos, 
after which $j_{\rm DM}$ is expected to be conserved \citep{Catelan96a}. 
Before turnaround halos continue to acquire angular momentum {as they grow in mass}. 
Turnaround epochs {of the $z=0$ host halos} are shown in Fig.~\ref{JVSMsR50Evo}  
as vertical segments. 
On the other hand, the half-mass radius of the stellar component 
 grows by $\gtrsim 0.7-0.9$~dex over the same period of time {and at fixed stellar mass}. 
This is interesting since in Fig.~\ref{JVSMsR50Evo} we focus on $j$ measured within $r_{\rm 50}$, 
{which implies that the radial profiles of $j_{\rm stars}$ in galaxies grow inside out. By studying the cumulative radial 
profiles of $j_{\rm stars}$ of the galaxies in Fig.~\ref{JVSMsR50Evo} we find that typically galaxies 
have profiles becoming steeper with decreasing redshift, and that at $z\gtrsim 1$ the inner regions of galaxies 
evolve very weakly, while the outer regions display a fast increase of $j_{\rm stars}$ (not shown here). 
These trends result in $j_{\rm stars}(r_{\rm 50})$ not evolving 
or only slightly increasing (in the case of star-forming galaxies) at $z<1$, even though 
$j_{\rm stars}(r)$, with $r\lesssim 6$~pkpc, decreases in the same period of time. The former is therefore 
a consequence of $r_{\rm 50}$ rapidly increasing with time.}

\citet{Catelan96a} predicted from linear tidal torque theory in a CDM universe 
that a halo collapsing at turnaround has an angular momentum of $L\propto M^{5/3}\,t^{1/3}$, where 
the time dependence comes from how the collapse time depends on halo mass, and thus at fixed halo mass, 
$j_{\rm halo}\propto M^{2/3}\,t^{1/3}$ at the moment of collapse. 
\citet{Catelan96a} also showed that in an Einstein de Sitter universe, 
the angular momentum of material falling into halos has $L\propto t$, which means that material falling 
later brings higher $L$. Under $j$ conservation, one could assume that $j_{\rm stars}$ follows a similar 
behaviour. We show in Fig.~\ref{JVSMsR50Evo}, using an arbitrary normalisation, how these time scalings 
compare with the evolution of \eagle\ galaxies. 
{$j_{\rm stars}(r_{\rm 50})$ closely follows the scaling of $t^{1/3}$ before the turnaround epoch 
while after turnaround $j_{\rm stars}(r_{\rm 50})$ is mostly flat (except for massive star-forming galaxies, that continue to display 
$j_{\rm stars}(r_{\rm 50})$ increasing), while $j_{\rm neutral}(r_{\rm 50})$ evolves close to $\propto t$. The latter is expected 
if the neutral gas is being freshly supplied by gas that is falling into halos.}
{The comparison with the expected time scalings of \citet{Catelan96a} should be taken as reference only, given that here 
we are not tracing the progenitors of galaxies, and thus the evolution seen in Fig.~\ref{JVSMsR50Evo} does not correspond to individual galaxies. 
In $\S$~\ref{TraceBackj} we study how $j_{\rm stars}$ developed in individual galaxies, selected at $z=0$.}

\subsection{Tracing the development of $j$ in individual galaxies}\label{TraceBackj}

Until now we have studied the evolution of $j$ at fixed mass throughout time, but mass is also a dynamic property, and thus 
in the $j$-$M$ plane both quantities are evolving in time. To quantify how much $j$ changes in a given galaxy, 
we look at all galaxies with $M_{\rm stars}>10^9\,\rm M_{\odot}$ at $z=0$ and trace back their progenitors. 
By doing so we keep the mass axis fixed (at $z=0$).
We show in Fig.~\ref{SigmaEvoAscendant} the growth of $j_{\rm stars}$ 
at fixed $M_{\rm stars}$ at $z=0$. $j_{\rm stars}$ is measured within $r_{\rm 50}$ at different redshifts.
Galaxies with stellar masses $<10^{10}\,\rm M_{\odot}$ at $z=0$ 
gain most of their $z=0$ $j_{\rm stars}(r_{\rm 50})$ at $z<1$. {Between 
$10^{10}\,\rm M_{\odot}<M_{\rm stars}<10^{10.7}\,\rm M_{\odot}$ there is a transition, 
in a way that galaxies with $M_{\rm stars}>10^{10.7}\,\rm M_{\odot}$ at $z=0$ show the opposite 
behaviour}, with most of their $j_{\rm stars}$ having been acquired at $z\gtrsim 1$.
The latter display a rapid growth of $j_{\rm stars}$ at $1.2<z<3$ of $\approx 0.3-0.5$~dex, followed by a much 
slower growth at $0.5<z<1.2$ of $\approx 0.15$~dex.
At $z<0.5$ these massive galaxies have $j_{\rm stars}(r_{\rm 50})$ even decreasing, due to the incidence of 
dry mergers (those with $f_{\rm gas,neutral}\lesssim 0.1$; Lagos et al. in preparation). 
Galaxies with stellar masses at $z=0$ in the range
$10^{10.1}-10^{10.7}\,\rm M_{\odot}$ are the ones experiencing the largest increase in $j_{\rm stars}$ (Fig.~\ref{SigmaEvoAscendant}).
These galaxies grow their $j_{\rm stars}$ by $\approx 0.7-1$~dex from $\approx 3$ to $z=0$. 
We find that $j_{\rm bar}(r_{50})$ evolves very similarly to what is shown in Fig.~\ref{SigmaEvoAscendant}.

{Fig.~\ref{RadProfilesJstars} shows the evolution of the cumulative profiles of $j_{\rm stars}$ for galaxies selected by 
their $z=0$ stellar mass, in the redshift range $0\le z \le 3$. 
We find that $j_{\rm stars}$ in the inner regions of galaxies evolves faster than in the outer regions. This 
is particularly dramatic at the highest stellar mass bin shown in Fig.~\ref{RadProfilesJstars}, where 
the total $j_{\rm stars}$ (measured with all the star particles of the sub-halo) 
{increases by $\approx 0.5$~dex from $z\approx 3$ to $z\approx 0.8$, followed by a decline of $\approx 0.1$ down to $z=0$,} 
while within $r_{50}$, $j_{\rm stars}$ increases by $\approx 1$~dex 
from $z\approx 3$ to $z\approx 0.8$, followed by a decrease {of $\approx 0.2$~dex} down to $z=0$. 
In the smallest mass bin the effect is subtle and there 
is only $\approx 0.1$~dex difference between the evolution of $j_{\rm stars}(r_{50})$ and $j_{\rm stars}(\rm tot)$ 
at $0\le z \le 3$, while at $z\lesssim 0.8$ the inner $j_{\rm stars}$ 
increases faster than the total $j_{\rm stars}$ by a factor of $\approx 1.4$. The latter effect is even stronger when 
young galaxies are considered (Fig.~\ref{SigmaEvoAscendantAgeProperties}). We will come back to 
this point in $\S$~\ref{tracks}. {The effect described here is partially due to $r_{50}$ increasing with time, which 
causes $j_{\rm stars}(r_{50})$ to also increase, but also due to an evolution in how $j_{\rm stars}$ is radially 
distributed in galaxies.} 

\eagle\ shows that in addition to the total $j_{\rm stars}$ of galaxies evolving, they also suffer from significant radial rearrangement 
of their $j_{\rm stars}$ throughout their lifetimes.}

\subsubsection{Evolutionary tracks of $j/M^{2/3}$}\label{tracks}

\begin{table}
\begin{center}
\caption{{Average evolutionary tracks of $\lambda^{\prime}_{\rm stars}\equiv j_{\rm stars}(r_{50})/M^{2/3}_{\rm stars}$ 
for galaxies that never had a merger, or those that have had at least $1$ merger, and divided into 
galaxies with mass-weighted stellar ages $>$ or $<$ $9$~Gyr. The units of 
 $\lambda^{\prime}_{\rm stars}$ are $\rm pkpc\,km\,s^{-1}$. These evolutionary tracks are shown as dotted lines 
in Fig.~\ref{SigmaEvoAscendantAge}. We also show in parenthesis the percentage of $z=0$ galaxies with 
$M_{\rm stars}(z=0)\ge 10^{9.5}\,\rm M_{\odot}$ and $r_{50}(z=0)>1$~pkpc in \eagle\ 
that roughly follow each evolutionary path.}}\label{EvoTracksAve}
\begin{tabular}{l l }
\\[3pt]
\hline
                                               & No mergers              \\
\hline
$\langle\rm age_{\rm stars}\rangle\ge  9\,Gyr$ & $\lambda^{\prime}_{\rm stars} =10^{-4.9}$ \\
($\approx 2$\%)                                  &                                  \\
\hline
$\langle\rm age_{\rm stars}\rangle<  9\,Gyr$   & $\lambda^{\prime}_{\rm stars} = 10^{-4.9},\,\,\,\,\,\,\,$ if $z\ge 1.2$ \\
($\approx 10$\%)                                 & $\,\,\,\,\,\,\,\,\,\,\,\,\,\,\,\,\,\,\,\,\,\,\,10^{-4.55}\,a,$ if $z<1.2$             \\
\hline
\hline
                                               & $N_{\rm mergers}>0$\\
\hline
$\langle\rm age_{\rm stars}\rangle\ge  9\,Gyr$ & $\lambda^{\prime}_{\rm stars} = 10^{-4.9},\,\,\,\,\,\,\,\,\,\,\,\,\,\,\,\,\,$ if $z\ge 1.2$\\
($\approx 47$\%)                                 & $\,\,\,\,\,\,\,\,\,\,\,\,\,\,\,\,\,\,\,\,\,\,\,10^{-5.15}a^{-0.7},$ if $z<1.2$\\
\hline
$\langle\rm age_{\rm stars}\rangle<  9\,Gyr$   & $\lambda^{\prime}_{\rm stars} = 10^{-4.85},\,\,\,\,\,\,\,\,\,$ if $z\ge 1.2$\\
($\approx 41$\%)                                 & $\,\,\,\,\,\,\,\,\,\,\,\,\,\,\,\,\,\,\,\,\,\,\,10^{-4.7}a^{0.4},$ if $z<1.2$ \\
\hline
\end{tabular}
\end{center}
\end{table}

There are two dominant effects that determine the value of $j_{\rm stars}$ at any one time in a galaxy's history: 
(i) whether stars formed before turnaround or after; those formed before tend to have lower $j_{\rm stars}$ than those formed after,
and (ii) whether galaxies have undergone dry galaxy mergers; 
these systematically lower $j_{\rm stars}$ in galaxies. 
We define the spin parameter of the stars, $\lambda^{\prime}_{\rm stars}\equiv j_{\rm stars}(r_{50})/M^{2/3}_{\rm stars}$ (as 
on average $j_{\rm stars}(r_{50})\propto M^{2/3}_{\rm stars}$ in \eagle), 
and show the evolution of $\lambda^{\prime}_{\rm stars}$ 
for 
galaxies that have different mass-weighted stellar ages at $z=0$ in Fig.~\ref{SigmaEvoAscendantAge}.
We name this parameter as $\lambda^{\prime}_{\rm stars}$ to distinguish it from the dimensionless 
spin parameter, defined in $\S$~\ref{theoryback}.
We separate galaxies that never suffered a galaxy merger (top panel) 
from those that went through at least one galaxy merger (bottom panel).
Here galaxy mergers are defined as those with 
a mass ratio $\ge 0.1$, while lower mass ratios are considered to be accretion \citep{Crain16}.

The top panel of Fig.~\ref{SigmaEvoAscendantAge} shows that 
galaxies with $\rm \langle\rm age_{\rm stars}\rangle\gtrsim 9\,Gyr$ have roughly 
constant $\lambda^{\prime}_{\rm stars}$ over time, albeit with 
 large scatter. Most of the stars in these galaxies were formed before the epoch of turnaround. On the other hand, 
galaxies with $\rm \langle\rm age_{\rm stars}\rangle\lesssim 9\,Gyr$ show a significant increase 
in their $\lambda^{\prime}_{\rm stars}$
at {$z\lesssim 1.2$}, after turnaround. The extent to which the latter galaxies increase 
their $\lambda^{\prime}_{\rm stars}$ is very similar, despite the wide spread in ages. 

In the subsample of galaxies that had at least one galaxy merger during their formation history (bottom panel of 
Fig.~\ref{SigmaEvoAscendantAge}), the effects of mergers are apparent. Galaxies with 
 $\rm \langle\rm age_{\rm stars}\rangle \gtrsim 9\,Gyr$ show a significant reduction of their 
$\lambda^{\prime}_{\rm stars}$
at {$z\lesssim 1.2$} where most of the mergers are dry.
Galaxies with $\rm \langle\rm age_{\rm stars}\rangle\lesssim 9\,Gyr$ that had mergers still show an increase of their 
 $\lambda^{\prime}_{\rm stars}$ at {$z\lesssim 1.2$} but to a lesser degree than the sample without mergers. 

From Fig.~\ref{SigmaEvoAscendantAge}, we extract average evolutionary tracks of 
$\lambda^{\prime}_{\rm stars}$. {These are presented in Table~\ref{EvoTracksAve}, along with the 
percentage of $z=0$ galaxies 
that followed each evolutionary path, and are shown as 
dotted lines in Fig.~\ref{SigmaEvoAscendantAge}.}
A powerful conclusion of Fig.~\ref{SigmaEvoAscendantAge} {and Table~\ref{EvoTracksAve}} is 
that {\it galaxies can have low $j_{\rm stars}$ either by the effects of mergers 
or by simply having formed most of their stars early on.} The simple picture from \citet{Fall83} invoked only mergers 
to explain the low $j_{\rm stars}$ of early-type galaxies. Here we find a more varied scenario.
{The latter statement holds regardless of the aperture used to measure $j_{\rm stars}$, however, the 
exact evolutionary tracks obtained are sensitive to the aperture used, as we describe below.} 

Fig.~\ref{tracksexample} shows examples
of these average tracks compare to the evolution of $\lambda^{\prime}_{\rm stars}$
in galaxies selected in bins of their host halo mass, and show that the they describe their {evolution relatively well and that the 
variations with halo mass are mild.}

{The tracks we identified in \eagle\ are partially driven by $j_{\rm stars}(r_{50})$ evolving more dramatically than the total
$j_{\rm stars}$ in galaxies. This is shown in Fig.~\ref{SigmaEvoAscendantAgeProperties} where the evolution of 
 $j_{\rm stars}(r_{50})$, $j_{\rm stars}\rm (tot)$ and $M_{\rm stars}$ are shown for galaxies in two bins of 
$\rm \langle\rm age_{\rm stars}\rangle$. The selection in $\rm \langle\rm age_{\rm stars}\rangle$ yields to two clear bins in stellar mass, which 
is due to the positive relation between $\rm \langle\rm age_{\rm stars}\rangle$ and stellar mass. In the case of 
old, massive galaxies, we find that before turnaround ($z\approx 1.2$ for these galaxies) $j_{\rm stars}\rm (tot)$ increases 
approximately as $\propto t^{1/3}$, consistent with the theoretical expectations of \citet{Catelan96a} discussed in $\S$~\ref{evojsec}, 
while in the same period of time $j_{\rm stars}(r_{50})$ increases faster. After turnaround, $j_{\rm stars}\rm (tot)$ 
shows very little evolution, while $j_{\rm stars}(r_{50})$ decreases by $\approx 0.3$~dex due to the effect of 
 galaxy mergers (Lagos et al. in preparation). 
On the other hand, younger, low-mass galaxies, have $j_{\rm stars}(r_{50})$ increasing very rapidly after turnaround, while 
$j_{\rm stars}\rm (tot)$ mostly grows before turnaround, and flattens after. The latter trends influence the evolutionary tracks 
of $\lambda^{\prime}_{\rm stars}$ presented in Table~\ref{EvoTracksAve}; i.e. the power-law indices change if we instead examine 
$j_{\rm stars}\rm (tot)$. Nonetheless, given that good quality kinematics is mostly available for the inner 
regions of galaxies, we consider the tracks presented here useful to test the predictions of \eagle. 
In addition, $j_{\rm stars}$ converges to $j_{\rm stars}\rm (tot)$ at $\approx 5 r_{50}$,
implying that good quality kinematic information is required up to that radii to carry out reliable 
measurements of $j_{\rm stars}\rm (tot)$.}

{The evolutionary tracks described here are connected to the variety of formation mechanisms of slow rotators in \eagle. 
In \eagle\ we find that $\approx 13$\% of galaxies in the mass range $10^{9.5}\,\rm M_{\odot}<M_{\rm stars}\lesssim 10^{10}\rm M_{\odot}$ 
at $z=0$ that have not suffered galaxy mergers 
 have $\lambda_{\rm R}\lesssim 0.2$. This percentage increases to $35$\% in galaxies of the same stellar masses but that had had 
mergers. Note, however, that galaxies that are slow rotators in \eagle\ and that never had a merger 
have exclusively low stellar masses, $M_{\rm stars}\lesssim 10^{10}\rm M_{\odot}$. 

The results presented here open up more complex formation paths of slow rotators than it has been suggested in the literature 
(e.g. \citealt{Emsellem11}), which has been mostly focused on galaxy mergers as the preferred formation scenario. \citet{Naab14} showed 
in $44$ simulated galaxies that this was indeed possible (see also \citealt{Feldmann11b}). 
Here we confirm this result with a much more statistically significant sample.}

\subsection{Comparison with theoretical models}\label{CompToModels}

\begin{figure}
\begin{center}
\includegraphics[width=0.49\textwidth]{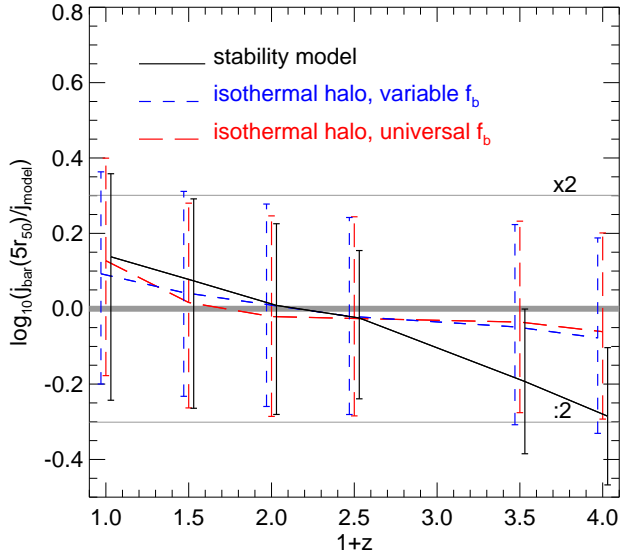}
\caption{The ratio between $j_{\rm bar}$, measured within $5r_{50}$, 
and the prediction of the isothermal sphere model assuming a Universal baryon fraction (Eq.~\ref{jhaloisoexample}; long-dashed line), 
and using the baryon fraction calculated for the individual subhalos where galaxies reside (Eq.~\ref{jhaloiso}; short-dashed line), for 
galaxies in \eagle\ with $M_{\rm stellar}>10^{9.5}\,\rm M_{\odot}$ (above the resolution limit; see Fig.~\ref{Convergence}) and $r_{50}>1$~pkpc. 
Also shown is the ratio between {$j_{\rm bar}(5r_{50})$}
and the predictions of the stability model of {\citet{Obreschkow16} (Eq.~\ref{stabilitymodel2}; solid line).}
Lines show the median, while the error bars show the $16^{\rm th}$ to $84^{\rm th}$ percentiles. 
{For reference, the horizontal thick solid line shows identity, while $\times 2$ above and $:2$ below identity are shown as horizontal thin solid lines}.}
\label{Residualsr50}
\end{center}
\end{figure}

In $\S$~\ref{theoryback} we introduced the expectations of two theoretical models, the isothermal collapsing halo 
with zero angular momentum losses, and the marginally stable disk model. Here we compare those expectations with our findings in \eagle.

{First, we use the {\tt HYPER-FIT} R package of \citet{Robotham15} to find the best fit between the properties 
$j_{\rm b}$, $\lambda_{\rm R}$ and $M_{\rm bar}$, with the former two being measured within $5\,r_{\rm 50}$. 
We find the best fit to be:} 

\begin{eqnarray}
\frac{j_{\rm b}}{\rm pkpc\,km\,s^{-1}} &\approx & 2.1\times10^{-5}\lambda^{1.08}_{\rm R}\,\left(\frac{M_{\rm bar}}{\rm M_{\odot}}\right)^{0.77}.
\label{jbarcalc}
\end{eqnarray}

\noindent {We can compare this fit with Eqs.~\ref{jhaloiso}~and~\ref{jhaloisoexample}, which correspond to the 
prediction of the isothermal collapsing halo
 with a varying baryon fraction and a Universal one, respectively.}
We can see that the best fit of Eq.~\ref{jbarcalc} is similar to the function of Eq.~\ref{jhaloiso}, with the best fit of \eagle\ 
having a slightly stronger dependency on both $\lambda_{\rm R}$ and $M_{\rm bar}$. The result of the isothermal collapsing halo 
model is compared to the true $j_{\rm bar}$ value of \eagle\ galaxies in 
Fig.~\ref{Residualsr50}, as long-dashed (Universal $f_{\rm b}$) and short-dashed lines (varying $f_{\rm b}$).
 In the case of the Universal $f_{\rm b}$, we adopted the value of \citet{Planck14}, while in the case of 
varying $f_{\rm b}$, we use the one calculated for each subhalo, where $f_{\rm b}=(M_{\rm stars}+M_{\rm neutral})/M_{\rm tot}$, where 
$M_{\rm tot}$ is the total mass of the subhalo.
{This simple model gives an expectation for $j_{\rm bar}$ that can differ from the true $j_{\rm bar}$ by up to $\approx 50$\%, on average
 (i.e. deviations from equity are $\lesssim 0.18$~dex, although the $1\sigma$ scatter around the median can be as large as $0.3$~dex).
There is a clear trend in which the model overestimates the true $j_{\rm bar}$ at high redshift, and underestimates it at low redshift.
Despite this trend, the simple isothermal sphere model is surprisingly} successful given the many physical processes 
that are included in \eagle\ but not in 
the model. The implications of this result are indeed deep, since this means that to some extent the assumptions made in semi-analytic models 
to connect the growth of halos with that of galaxies \citep{White91,Cole00,Springel01} are not far from how the physics of galaxy formation works in 
highly sophisticated, non-linear simulations. 
\citet{Stevens16b} discuss 
how the assumptions made in semi-analytic models of galaxy formation fit within the results of \eagle.

We also studied in detail the subsample of galaxies with $\lambda_{\rm R}(r_{50})>0.6$ (rotationally supported galaxies) at $0\le z\le 3$ 
{to compare with the theoretical model of \citet{Obreschkow16} based on the stability of disks}. As expected, 
we find that the atomic gas fraction becomes an important property, so that the best fit of $j_{\rm bar}(5\,r_{50})$ becomes 

\begin{equation}
\frac{j_{\rm bar}(5\,r_{50})}{\rm pkpc\,km\,s^{-1}} \approx 7.23\times10^{-4}\,f_{\rm atom}^{0.44} \left(\frac{M_{\rm bar}}{\rm M_{\odot}}\right)^{0.6}.
\label{disksj}
\end{equation}

\noindent Here, the scatter perpendicular to the hyper plane is $\sigma_{\perp}=0.19$, while the scatter parallel to 
$j_{\rm bar}$ is $\sigma_{\|}=0.26$. In \eagle\ we find a much weaker dependence of 
$j_{\rm bar}(5\,r_{50})$ on both $f_{\rm atom})$ and $M_{\rm bar}$ compared to the theoretical 
expectation (Eq.~\ref{stabilitymodel2}). 
We compare the predictions of this model to $j_{\rm bar}$ of \eagle\ galaxies with $\lambda_{\rm R}(r_{50})>0.6$ in 
Fig.~\ref{Residualsr50}. {To do this, we require a measurement of the velocity dispersion of the gas 
in \eagle\ galaxies (Eq.~\ref{stabilitymodel2}). We measure the 1-dimensional velocity dispersion of the star-forming gas in \eagle, 
$\sigma_{\rm 1D,SF}$, using Eq.~\ref{Sigma1dstar} and all star-forming gas particles within $5r_{50}$}.
The model of \citet{Obreschkow16} describes reasonably well, within a factor of $\approx 1.5$, 
the evolution of $j_{\rm bar}$ in galaxies with $\lambda_{\rm R}>0.6$ { at $z\lesssim 2$}.
{At higher redshifts it significantly deviates from $j_{\rm bar}(5r_{50})$ of \eagle\ galaxies. 
There could be several causes. For example, the model assumes
thin, exponential disks, while \eagle\ galaxies have increasingly lower $V_{\rm rot}/\sigma_{\rm stars}$ with increasing redshift, and thus 
we do not expect them to be well described by thin disk models. In addition, 
 the dependence of $f_{\rm atom}$ on $j_{\rm bar}$ becomes weaker in the gas-rich regime, typical of high-redshift galaxies, and thus 
the gas fraction becomes an increasingly poorer predictor of $j_{\rm bar}$.}

\section{Conclusions}\label{ConcluSec}

We presented a comprehensive study of how $j$ of the stellar, baryon and neutral gas components of galaxies, depend on galaxy properties using 
the \eagle\ hydrodynamic simulation. Our main findings are:

\begin{itemize}
\item In the redshift range studied, $0\le z\le3$, galaxies having 
higher neutral gas fractions, lower stellar concentrations, younger stellar ages, 
bluer $\rm (u^*-r^*)$ colours and higher $V_{\rm rot}/\sigma_{\rm stars}$ have higher $j_{\rm stars}$ and 
$j_{\rm bar}$ overall. All the properties above are widely used 
 as proxies for the morphologies of galaxies, and thus we can comfortably conclude that late-type galaxies 
in \eagle\ have higher $j_{\rm stars}$ and $j_{\rm bar}$ than early-type galaxies, as observed.

\item We compare with $z=0$ observations and find that the trends seen in the $j$-mass plane reported by 
\citet{Romanowsky12}, \citet{Obreschkow14b}, \citet{Cortese16} and measured here for the ATLAS$^{\rm 3D}$ survey, 
with stellar concentration, 
neutral gas fraction and $\lambda_{\rm R}$, are all also present in \eagle\ in a way that resembles the observations very closely. 
These trends show that galaxies with lower $\lambda_{\rm R}$, lower gas fractions and higher stellar concentrations, generally have lower 
$j_{\rm stars}$ and $j_{\rm bar}$ at fixed stellar and baryon mass, respectively. Again, the trends above are present regardless of the 
apertures used to measure $j$.
 
\item $j$ scales with mass roughly as $j\propto M^{2/3}$ for both the stellar and total baryon components of galaxies. This is the case 
for all galaxies with $M_{\rm stars}>10^9\,\rm M_{\odot}$ at $0\le z\le 3$. In the case of the neutral gas we find a different scaling closer to 
$j_{\rm neutral}\propto M^{1/3}_{\rm neutral}$, which we attribute to the close relation between $j_{\rm neutral}$ and $j$ of the entire halo 
\citep{Zavala15} and the poor correlation between the neutral gas content of galaxies and the halo properties.

\item We identified two generic tracks for the evolution of the stellar spin parameter, 
$\lambda^{\prime}_{\rm stars}\equiv j_{\rm stars}(r_{50})/M^{2/3}_{\rm stars}$, 
 depending on whether most of stars formed before or after turnaround (which occurs at $z\approx 0.85$ for galaxies 
that at $z=0$ have $M_{\rm stars}>10^{9.5}\,\rm M_{\odot}$).
In the absence of mergers, 
galaxies older than $9\,\rm Gyr$ (i.e. most stars formed before turnaround) show little evolution in their $j_{\rm stars}/M^{2/3}_{\rm stars}$,
while younger ones show a constant $\lambda^{\prime}_{\rm stars}$ until $z\approx 1.2$, 
and then increase as $\lambda^{\prime}_{\rm stars}\propto a$. Mergers 
reduce $\lambda^{\prime}_{\rm stars}$ by factors of $\approx 2-3$, on average, in galaxies older than $9\,\rm Gyr$, 
and the index of the scaling between $\lambda^{\prime}_{\rm stars}$ and the scale factor to $\approx 0.4$ in younger galaxies.
{We find that these tracks are the result of two effects: (i) the evolution of the total $j_{\rm stars}$ of galaxies, and (ii) 
its radial distribution, which suffers significant rearrangements in the inner regions of galaxies at $z\lesssim 1$.}
{Regardless of the aperture in which $j_{\rm stars}$ is measured}, two distinct channels leading to low $j_{\rm stars}$ in galaxies at $z=0$
are identified: (i) galaxy mergers, and (ii) early formation of
most of the stars in a galaxy.

\item We explore the validity of 
two simple, theoretical models presented in the literature that follow the evolution of $j$ in galaxies using 
\eagle. We find that on average \eagle\ galaxies follow the predictions of an 
isothermal collapsing halo with negligible angular momentum losses within a factor of $\approx 2$.
These results are interesting, as it helps validating 
some of the assumptions that go into the semi-analytic modelling 
technique to determine $j$ and sizes of galaxies (e.g. \citealt{White91,Kauffmann93,Cole00}), at least as a net 
effect of the galaxy formation process. 
We also test the model of \citet{Obreschkow16},
in which the stability of disks is governed by the disk's angular momentum. In this model, 
$f_{\rm atom}\propto (j_{\rm bar}/M_{\rm bar})^{1.12}$. {We find that this model can reproduce the 
 evolution of $j_{\rm bar}$ to within $50$\% at $z\lesssim 2$}, but only of \eagle\ galaxies that are 
rotationally-supported. 
\end{itemize}

{One of the most important predictions that we presented here 
is the evolution of $j_{\rm stars}(r_{50})$ in passive and active galaxies, and 
the evolutionary tracks of $\lambda^{\prime}_{\rm stars}$.}
The advent of high quality IFS
instruments and experiments such as the SKA, discussed in $\S$~$1$, will open 
the window to measure $j$ at redshifts higher than $0$, and to increase the number of galaxies with accurate 
measurements of $j$ by one to two orders of magnitude. They will be key to study the co-evolution of the quantities addressed here 
 and test our \eagle predictions.

\section*{Acknowledgements}

We thank Charlotte Welker, Danail Obreschkow, Dan Taranu, Alek Sokolowska, Lucio Mayer, Eric Emsellem 
and Edoardo Tescari for inspiring and useful discussions.
We also thank the anonymous referee for a very insightful report.  
CL is funded by a Discovery Early Career Researcher Award (DE150100618).
CL also thanks the MERAC Foundation for a Postdoctoral Research Award and the organisers of the ``Cold Universe'' KITP programme 
for the opportunity to attend and participate in such an inspiring workshop. 
This work was supported by a Research Collaboration Award 2016 at the University of Western Australia.
This work used the DiRAC Data Centric system at Durham University, operated by the Institute for Computational Cosmology on behalf of the STFC DiRAC HPC Facility ({\tt www.dirac.ac.uk}). This equipment was funded by BIS National E-infrastructure capital grant ST/K00042X/1, STFC capital grant ST/H008519/1, and STFC DiRAC Operations grant ST/K003267/1 and Durham University. DiRAC is part of the National E-Infrastructure.
Support was also received via the Interuniversity Attraction Poles Programme initiated
by the Belgian Science Policy Office ([AP P7/08 CHARM]), the
National Science Foundation under Grant No. NSF PHY11-25915,
and the UK Science and Technology Facilities Council (grant numbers ST/F001166/1 and ST/I000976/1) via rolling and
consolidating grants awarded to the ICC.
We acknowledge the Virgo Consortium for making
their simulation data available. The \eagle\ simulations were performed using the DiRAC-2 facility at
Durham, managed by the ICC, and the PRACE facility Curie based in France at TGCC, CEA, Bruyeres-le-Chatel.
This research was supported in part by the National Science Foundation under Grant No. NSF PHY11-25915.
Parts of this research were conducted by the Australian Research Council Centre of Excellence for All-sky Astrophysics (CAASTRO), through project number CE110001020.

%----------------------------------------------
\bibliographystyle{mn2e_trunc8}
\bibliography{AM}
%---------------------------------------------------------------------

\label{lastpage}
\appendix
\section[]{Strong and weak convergence tests}\label{ConvTests}

\begin{table*}
\begin{center}
\caption{\eagle\ simulations used in this Appendix. The columns list:
    (1) the name of the simulation, (2) comoving box size, (3) number
    of particles, (4) initial particle masses of gas and (5) dark
    matter, (6) comoving gravitational
    softening length, and (7) maximum physical comoving Plummer-equivalent
    gravitational softening length. Units are indicated below the name of
    each column. \eagle\
    adopts (6) as the softening length at $z\ge 2.8$, and (7) at $z<2.8$. The simulation Recal-L025N0752 
    has the same
    masses of particles and softening length values than the simulation Ref-L025N0752.}\label{TableSimus2}
\begin{tabular}{l c c c c c l}
\\[3pt]
\hline
(1) & (2) & (3) & (4) & (5) & (6) & (7) \\
\hline
Name & $L$ & \# particles & gas particle mass & DM particle mass & Softening length & max. gravitational softening \\
Units & $[\rm cMpc]$   &                &  $[\rm M_{\odot}]$ &  $[\rm M_{\odot}]$ & $[\rm ckpc]$ & $[\rm pkpc]$\\
\hline
Ref-L025N0376 & $25$ &   ~$2\times 376^3$  &$1.81\times 10^6$   &   ~~$9.7\times 10^6$ & $2.66$ & ~~$0.7$  \\
Ref-L025N0752 & $25$ &   ~$2\times 752^3$  &$2.26\times 10^5$   &   $1.21\times 10^6$  & $1.33$ &  $0.35$ \\
\hline
\end{tabular}
\end{center}
\end{table*}

\begin{figure*}
\begin{center}
\includegraphics[width=0.4\textwidth]{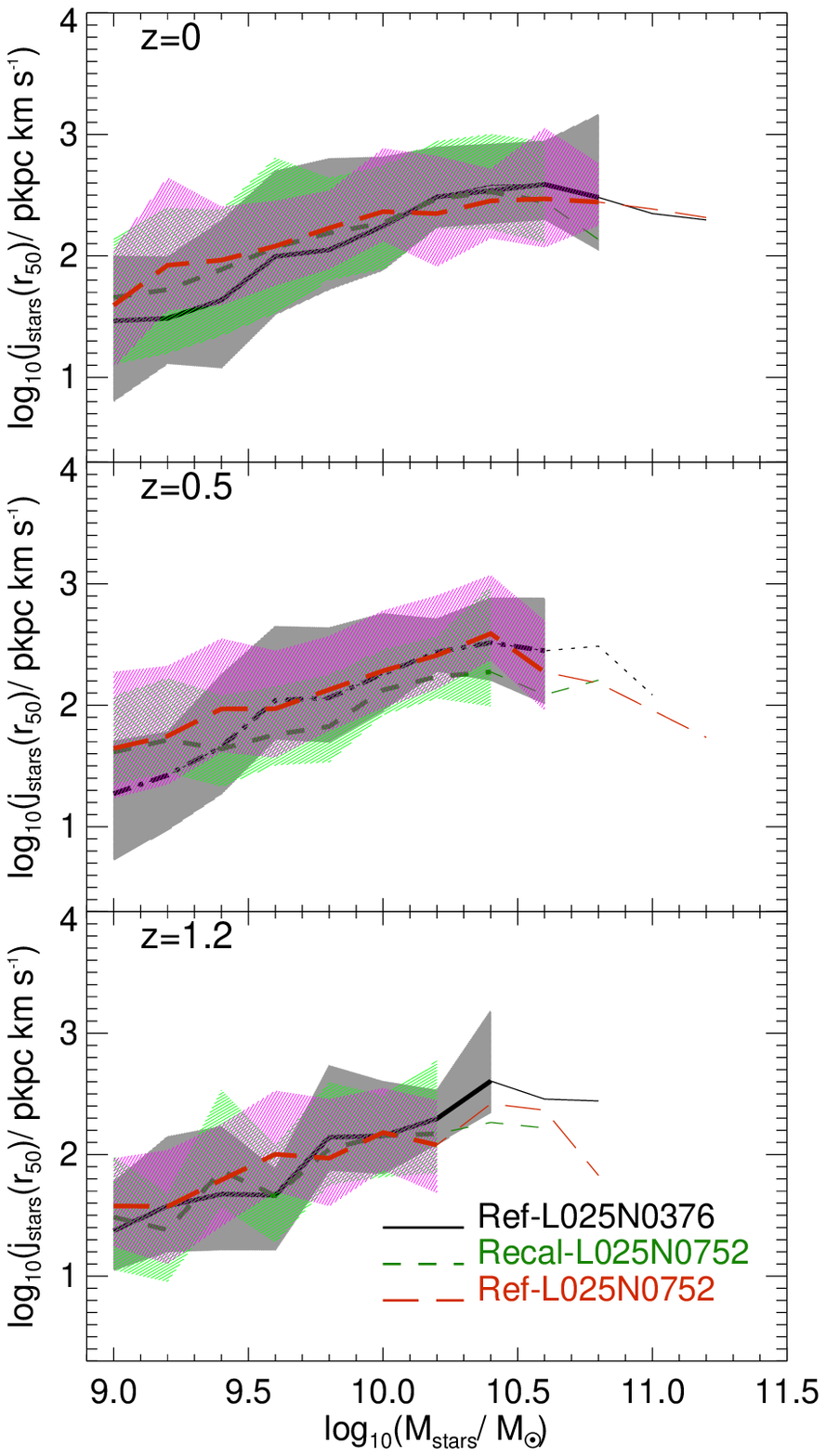}
\includegraphics[width=0.4\textwidth]{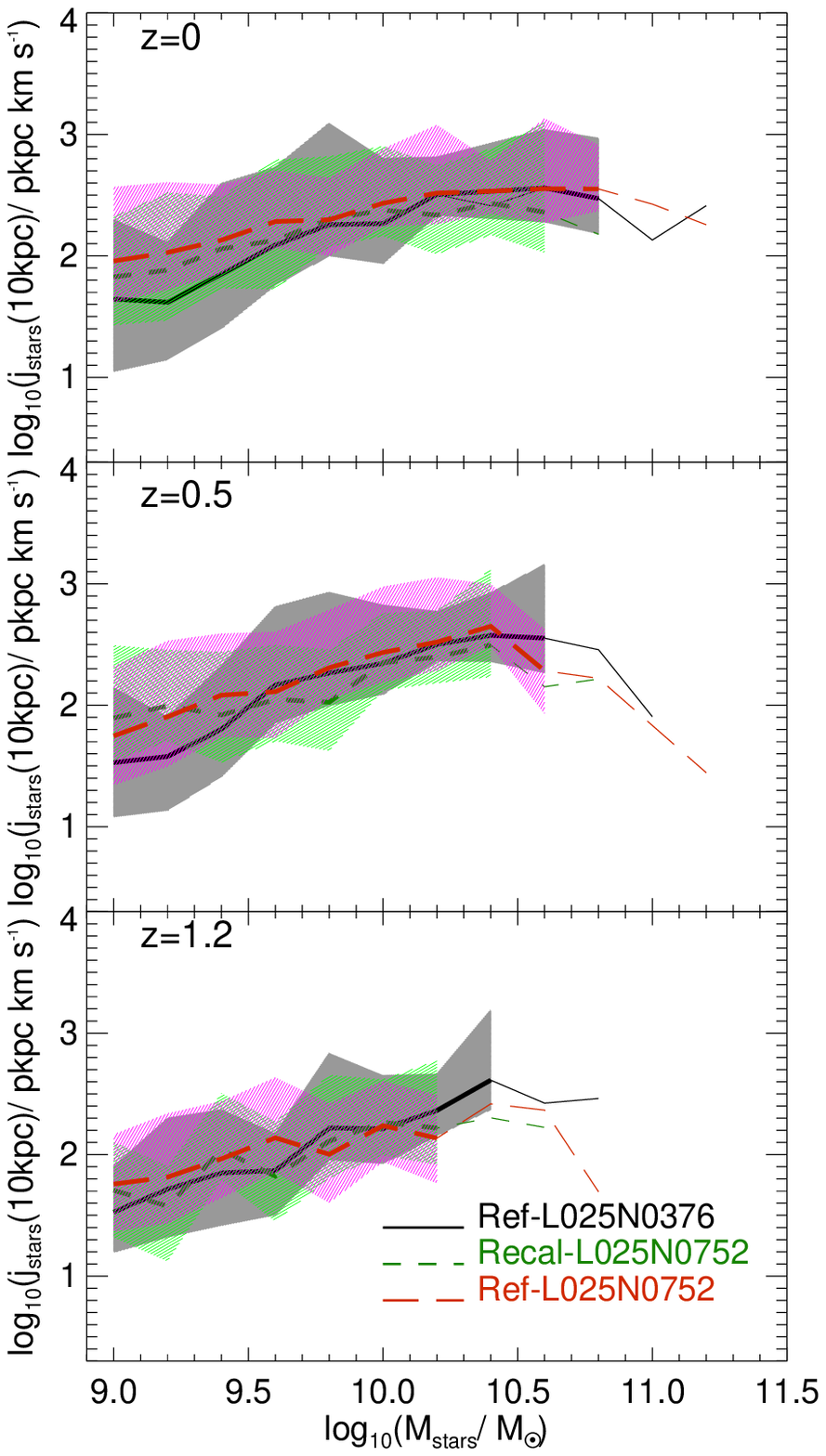}
\caption{The $j_{\rm stars}$-stellar mass relation at three redshifts, $z=0,\,0.5\,1.2$, for the 
 Ref-L025N0376, Ref-L025N0752 and Recal-L025N0752 simulations, as labelled. We show the relation with $j_{\rm stars}$ 
measured within the half mass radius of the stellar component (left panels) and within a fixed aperture of 
$10$~pkpc. Lines show the median relations, while the shaded regions show the
$16^{\rm th}-84^{\rm th}$ percentile ranges. The latter are presented only for bins with $\ge 10$ galaxies. Bins with 
fewer objects are shown as thin lines.}
\label{Convergence}
\end{center}
\end{figure*}

\begin{figure}
\begin{center}
\includegraphics[width=0.4\textwidth]{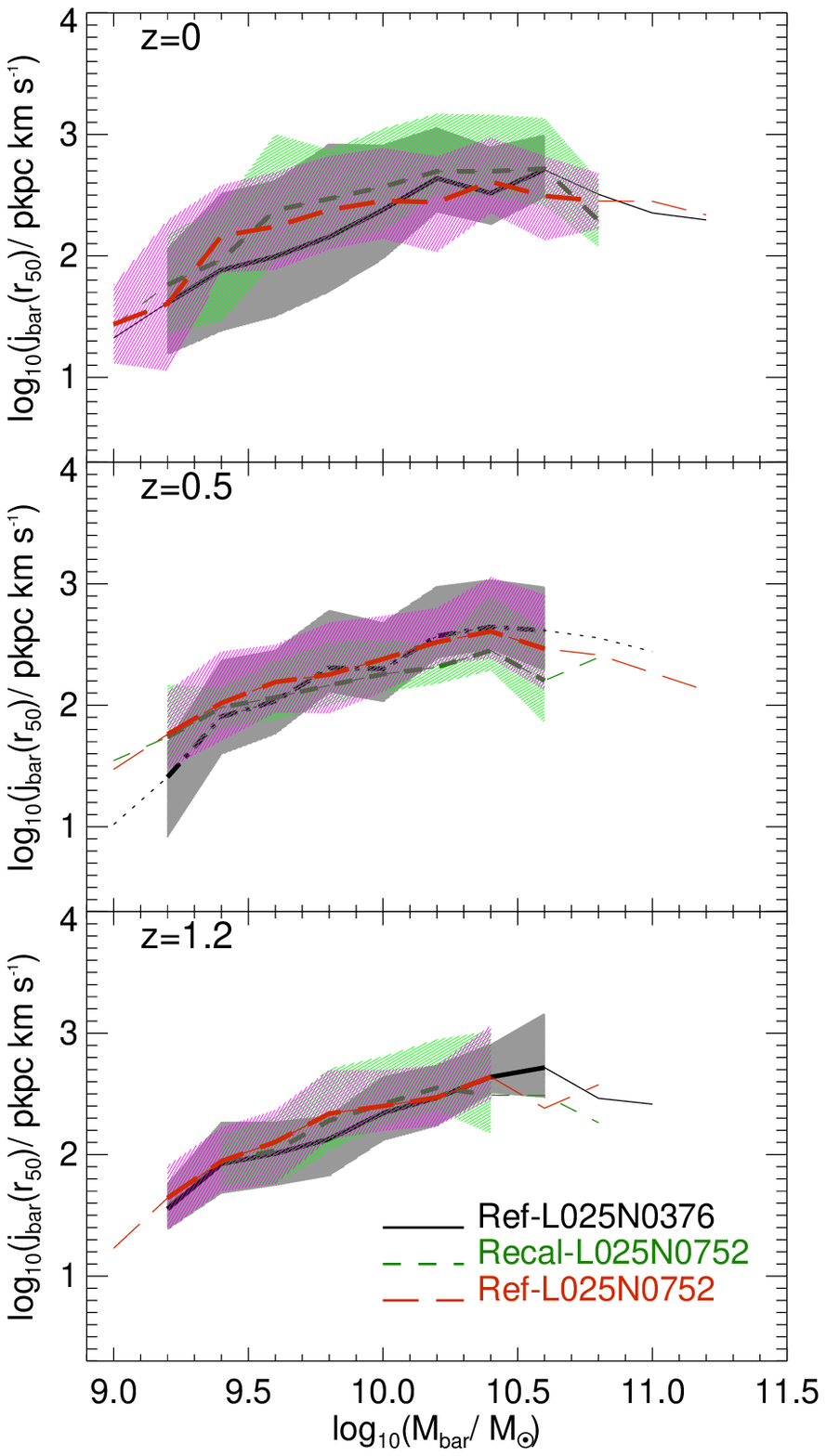}
\caption{As in the left panels of Fig.~\ref{Convergence} but here we show the $j_{\rm bar}$-baryon mass relation.}
\label{Convergence2}
\end{center}
\end{figure}

\begin{figure}
\begin{center}
\includegraphics[width=0.4\textwidth]{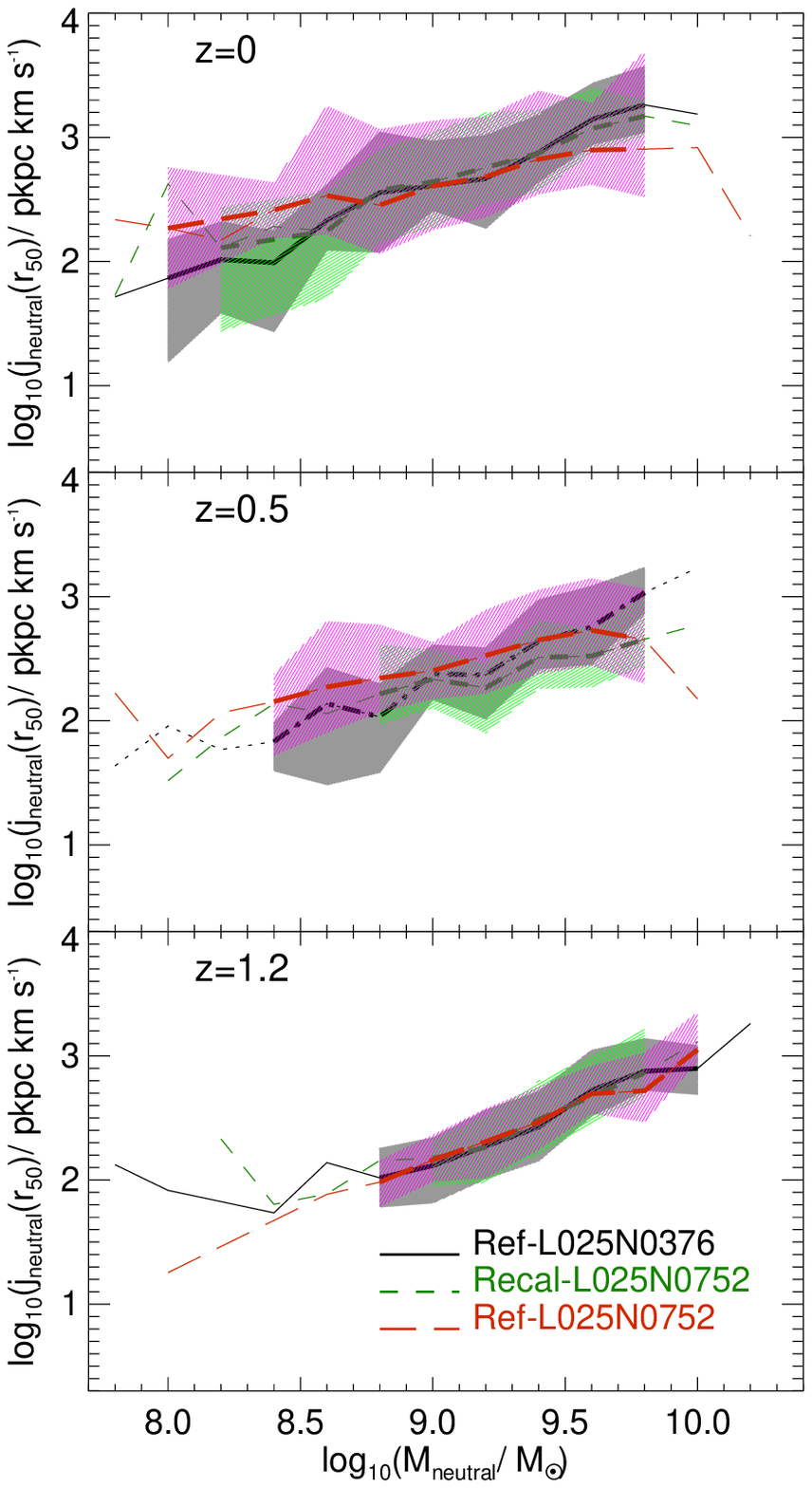}
\caption{As in the left panels of Fig.~\ref{Convergence} but here we show the $j_{\rm neutral}$-neutral gas mass relation.}
\label{Convergence3}
\end{center}
\end{figure}

S15 introduced the concept of `strong' and `weak' convergence
tests. Strong convergence refers to the case where a simulation is
re-run with higher resolution (i.e. better mass and spatial resolution)
adopting exactly the same subgrid physics and parameters. Weak
convergence refers to the case when a simulation is re-run with higher
resolution but the subgrid parameters are recalibrated to recover, as
far as possible, similar agreement with the adopted calibration
diagnostic (in the case of \eagle, the $z=0.1$ galaxy stellar mass
function and disk sizes of galaxies). 

S15 introduced two higher-resolution versions of \eagle, both in a box of
($25$~cMpc)$^{3}$ and with $2\times 752^3$ particles, Ref-L025N0752
and Recal-L025N0752 (Table~\ref{TableSimus2} shows some details of these simulations). 
These simulations have better mass and spatial
resolution than the intermediate-resolution simulations by factors of
$8$ and $2$, respectively. In the case of Ref-L025N0752, the parameters of the sub-grid physics are 
kept fixed (and therefore comparing with this simulation is a strong convergence test), while 
the simulation Recal-L025N0752 has $4$ parameters whose values have
been slightly modified with respect to the reference simulation (and therefore comparing with this simulation is a weak 
convergence test).

Here we compare the relation between $j_{\rm stars}$ and stellar mass at three different redshifts 
in the simulations Ref-L025N0376, Ref-L025N0752 and Recal-L025N0752. Fig.~\ref{Convergence} shows
the $j_{\rm stars}-M_{\rm stars}$ relation, with $j_{\rm stars}$ measured in two different ways: (i) with all the star particles 
within a half-mass radius of the stellar component (this is what we do throughout the paper; left panels), and (ii) 
with all the star particles at a fixed physical aperture of $10$~pkpc (right panels). For the measurement of $j_{\rm stars}(r_{50})$ 
we find that the simulations Ref-L025N0376 and Ref-L025N0752 produce a very similar relation in the three redshifts 
analysed (within $\approx 0.15$~dex), 
$z=0,\,0.5\,1.2$. On the other hand, the Recal-L025N0752 simulation produces a $j_{\rm stars}(r_{50})-M_{\rm stars}$ relation at $z=0$ 
in very good agreement, but that systematically deviates with redshift. 
We find that this is due to the difference in the predicted stellar mass-$r_{50}$ relation between the different simulations. This is clear 
from the right panels of Fig.~\ref{Convergence}, where we compare now the $j_{\rm stars}(10\rm\,pkpc)-M_{\rm stars}$ relation. 
Here we see that the three simulations 
are generally consistent throughout redshift. One could argue that the intermediate resolution run, 
Ref-L025N0376, which corresponds to the resolution we use 
throughout the paper, tends to produce $j_{\rm stars}(10\rm\,pkpc)$ 
slightly smaller than the higher resolutions runs Ref-L025N0752 and Recal-L025N075 at 
$M_{\rm stars}\lesssim 10^{9.5}\,\rm M_{\odot}$. However, the effect is not seen at every redshift we analysed, 
and thus it could be due to statistical variations (note that 
the offset is much smaller than the actual scatter around the median). In order to be conservative, we show in the figures of this paper 
the limit of $M_{\rm stars}=10^{9.5}\,\rm M_{\odot}$, above which we do not see any difference that could make us 
suspect resolution limitations.

In Figs.~\ref{Convergence2}~and~\ref{Convergence3} we study the convergence of the relation 
$j_{\rm bar}-M_{\rm bar}$ and $j_{\rm neutral}-M_{\rm neutral}$ and conclude that the former is converged 
at $M_{\rm bar}\gtrsim 10^{9.5}\,\rm M_{\odot}$, while the latter is converged at 
$M_{\rm neutral}\gtrsim 10^{8.5}\,\rm M_{\odot}$.

\section[]{Scaling relations between the angular momentum of galaxy components}\label{ScalingRelations}

Here we present additional scaling relation between $j_{\rm neutral}$, $j_{\rm stars}$, stellar mass and other galaxy properties.

\begin{figure}
\begin{center}
\includegraphics[width=0.43\textwidth]{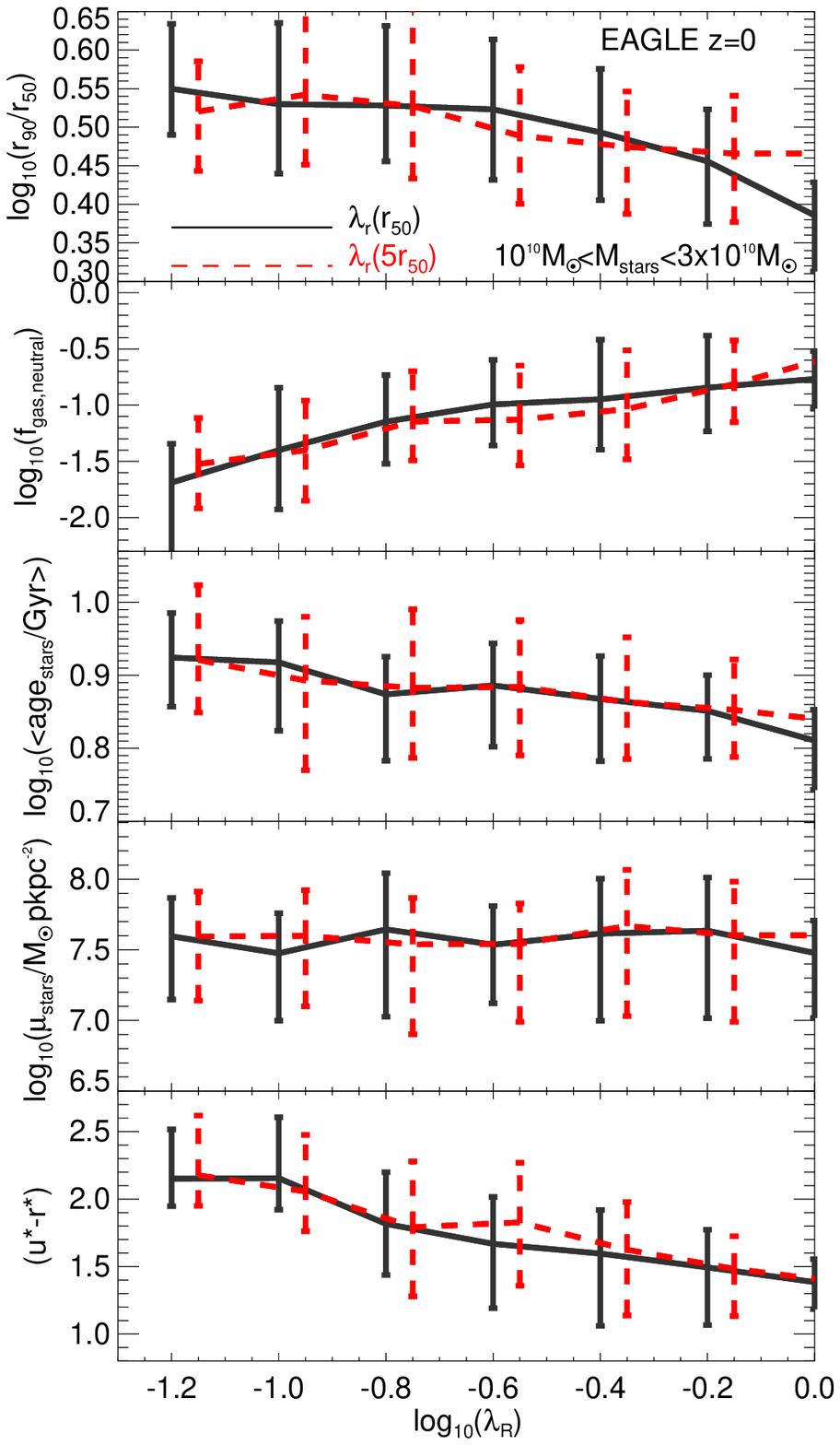}
\caption{The stellar concentration, $r_{\rm 50}/r_{90}$ (top panel),
neutral gas fraction (second panel), stellar age (middle panel),
central surface density of stars (fourth panel) and
$\rm (u^*-r^*)$ SDSS colour (bottom panel) as a function
of $\lambda_{\rm R}$, measured within $r_{50}$ (solid lines) and
$5\,r_{50}$ (dashed lines), for galaxies in \eagle\ at $z=0$ with $10^{10}\rm\,M_{\odot}<M_{\rm stars}<3\times 10^{10}\,\rm M_{\odot}$.
The lines show the medians with errorbars encompassing the $16^{\rm th}$ to $84^{\rm th}$ percentile ranges.}
\label{Morphology}
\end{center}
\end{figure}

\begin{figure}
\begin{center}
\includegraphics[width=0.49\textwidth]{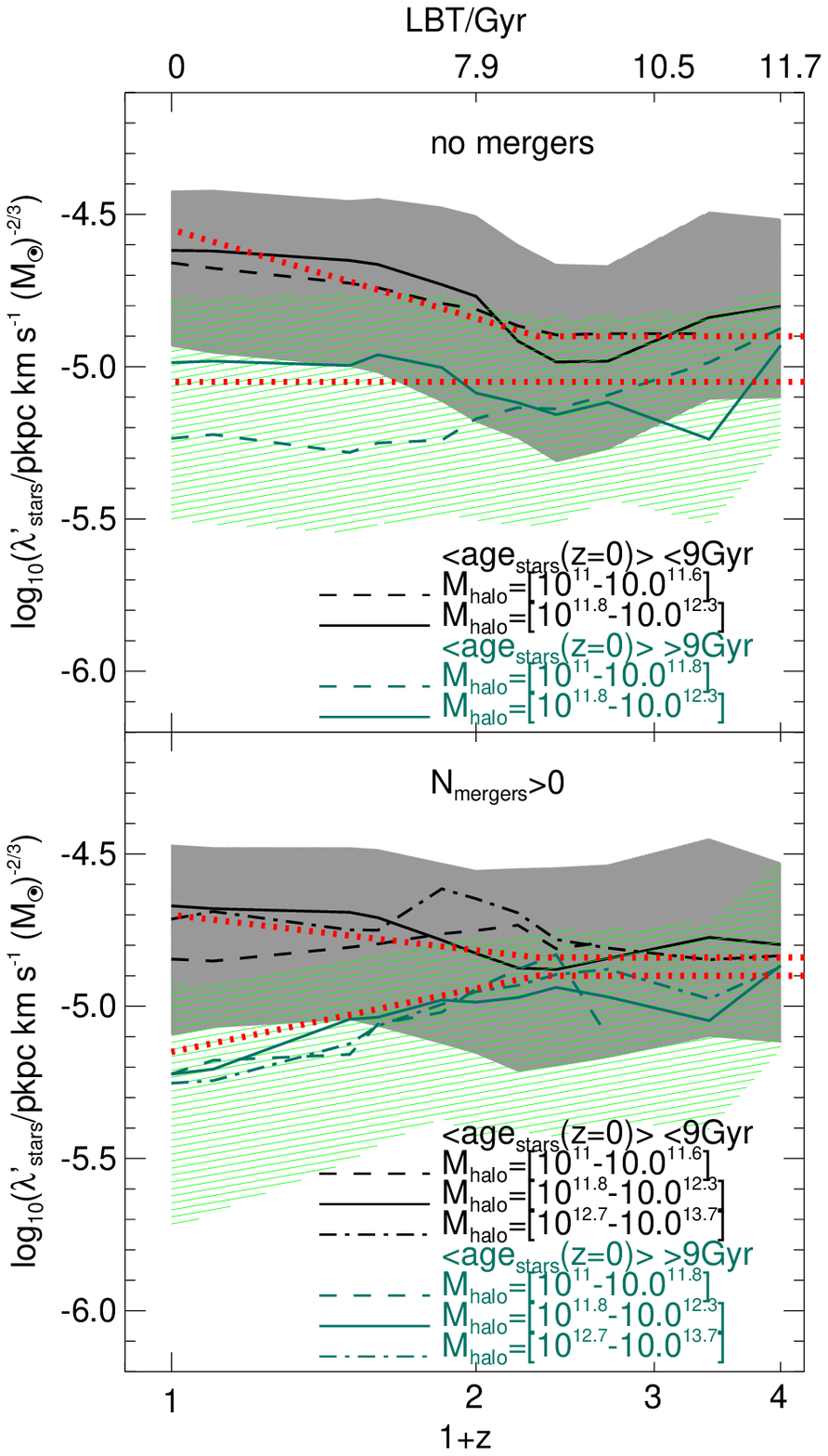}
\caption{The value of $\lambda^{\prime}_{\rm stars}\equiv j_{\rm stars}(r_{50})/M^{2/3}_{\rm stars}$ 
as a function of redshift in \eagle, for individual central galaxies
hosted by halos in the mass ranges labelled in the figure (in units of $\rm M_{\odot}$) at $z=0$,
and in two bins of $\langle\rm age_{\rm stars}\rangle$ at $z=0$, $\langle\rm age_{\rm stars}\rangle<9\,\rm Gyr$
(black lines) and $\langle\rm age_{\rm stars}\rangle>9\,\rm Gyr$ (green lines).
The top panel shows the subsample of the galaxies above that never suffered a galaxy merger, while in the bottom panel
we show those that had a least $1$ merger.
The shaded regions show $25^{\rm th}-75^{\rm th}$ percentile ranges, but only for the halo mass range
$10^{11.8}\,\rm M_{\odot}\le M_{\rm halo}\le 10^{12.3}\,\rm M_{\odot}$.
The evolutionary tracks of Fig.~\ref{SigmaEvoAscendantAge} are shown as dotted lines.}
\label{tracksexample}
\end{center}
\end{figure}

\begin{figure*}
\begin{center}
\includegraphics[trim=0.4cm 0cm 0cm 0cm, clip=true,width=0.98\textwidth]{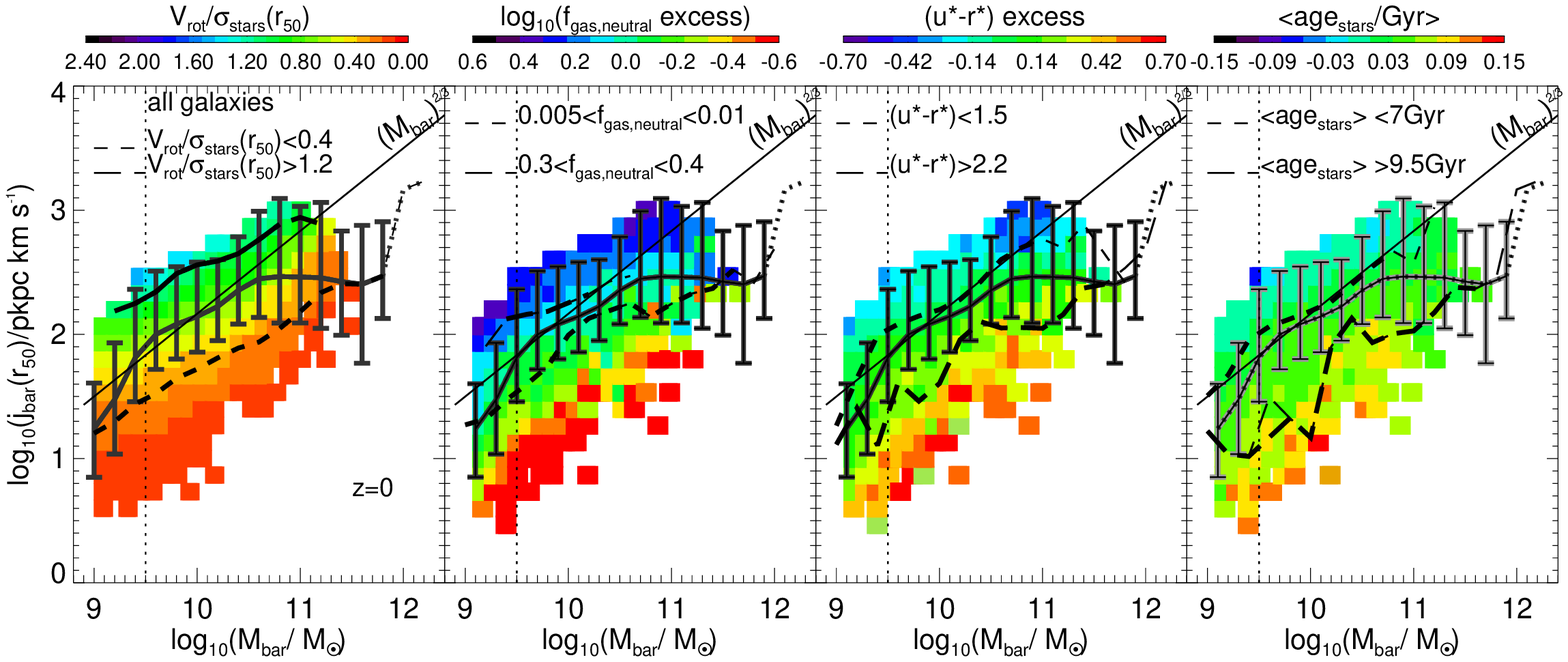}
\includegraphics[trim=0.4cm 0cm 0cm 0cm, clip=true,width=0.98\textwidth]{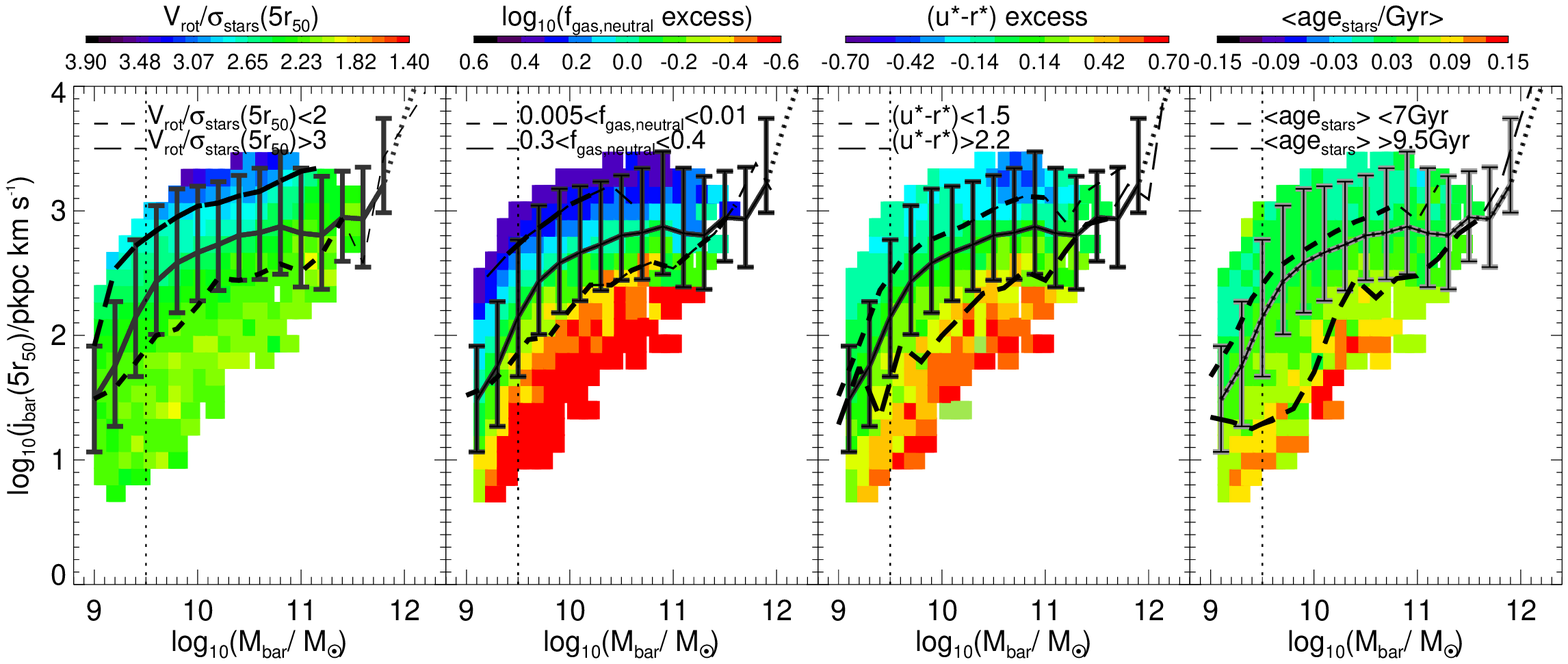}
\includegraphics[trim=0.4cm 0cm 0cm 0cm, clip=true,width=0.98\textwidth]{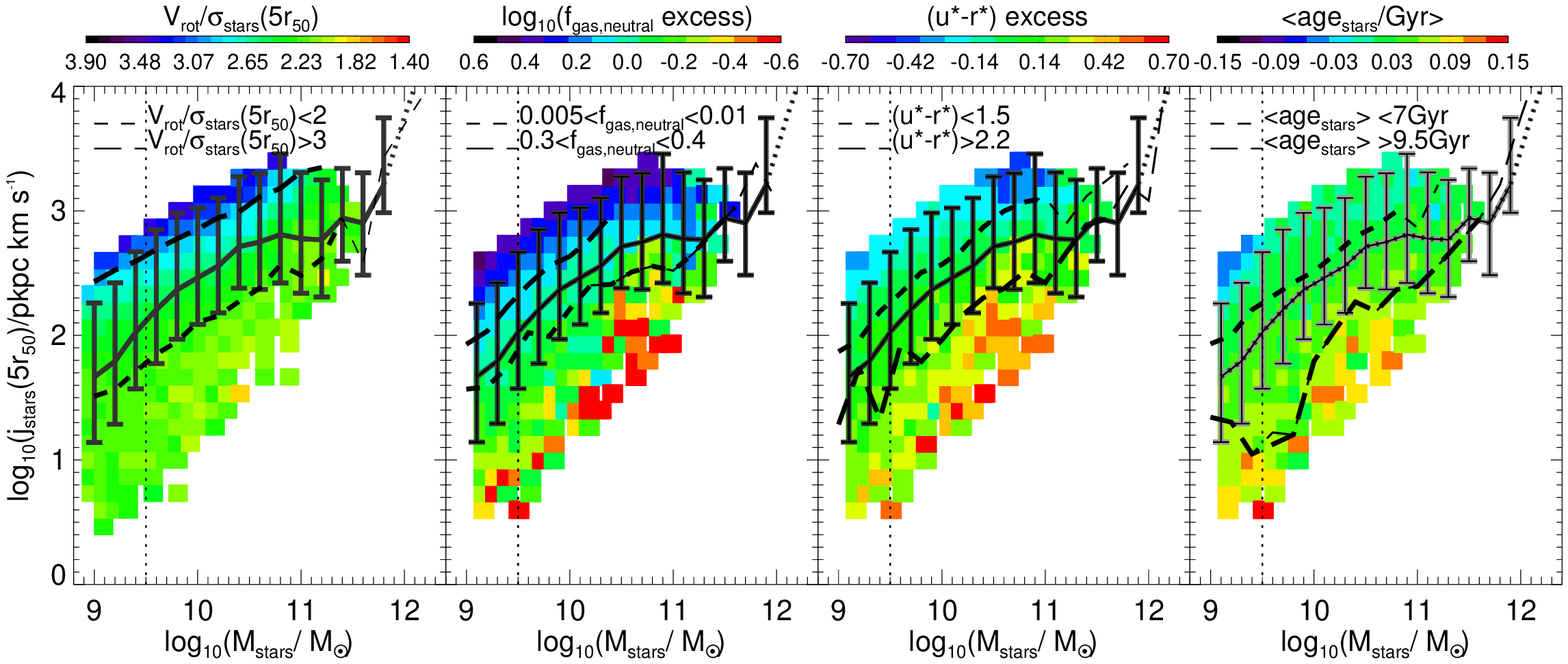}
\caption{{\it Top panels:} 
As Fig.~\ref{JMEAGLE} but for $j_{\rm bar}(r_{\rm 50})$ as a function of the baryon mass (stars plus neutral gas) at $z=0$.
{\it Middle panels:} As in the top panels but for $j_{\rm bar}(5r_{\rm 50})$.
{\it Bottom panels:} As Fig.~\ref{JMEAGLE} but for $j_{\rm stars}(5r_{\rm 50})$.}
\label{JBAREAGLE}
\end{center}
\end{figure*}

In \eagle\ we find that several galaxy properties that trace morphology are related to
$\lambda_{\rm R}$, which is used to define slow and fast rotators in the literature \citep{Emsellem07}. 
Fig.~\ref{Morphology} shows that at a given stellar mass, the neutral gas fraction,
the stellar concentration, stellar age, and $\rm (u^*-r^*)$ colour are correlated with $\lambda_{\rm R}$.
The latter is directly proportional to $j_{\rm stars}$ and thus it is expected that all these quantities correlate
with $j_{\rm stars}$. We do not find a relation between $\mu_{\rm stars}$ and
$\lambda_{\rm R}$, and indeed $\mu_{\rm stars}$ is poorly correlated with the positions
of galaxies in the $j_{\rm stars}$-stellar mass plane. 

We test how much the average evolutionary tracks identified in $\S$~\ref{tracks} are mass-independent by replicating the
experiment of Fig.~\ref{SigmaEvoAscendantAge} but for galaxies in bins of halo mass.
In Fig.~\ref{tracksexample}, we show the evolution of $\lambda^{\prime}_{\rm stars}$
for galaxies hosted in halos of different mass ranges $z=0$, separated into galaxies
that never suffered a merger (top panel), and that has at least one merger (bottom panel).
In each panel we show the subsamples with a mass-weighted stellar age $\rm \langle age_{\rm stars}\rangle \le 9\,\rm Gyr$
(solid line) and $\rm \langle age_{\rm stars}\rangle > 9\,\rm Gyr$ (dashed line). In addition, as dotted lines we show
the average evolutionary tracks found in $\S$~\ref{tracks}. We find that, although some trends can be noisy,
these tracks are a reasonable description of the average behaviour observed for the different halo mass bins.

In the top panels of Fig.~\ref{JBAREAGLE} we study the $j_{\rm bar}-M_{\rm bar}$ relation, 
and how the scatter correlates with  $V_{\rm rot}/\sigma_{\rm stars}$, $f_{\rm gas,neutral}$, 
$\rm (u^*-r^*)$ and mass-weighted stellar age. 
We find that there is a positive correlation between $j_{\rm bar}$ and $M_{\rm bar}$ 
at $10^{9.5}\,\rm M_{\odot}\lesssim M_{\rm bar}\lesssim 10^{10.7}\,\rm M_{\odot}$, 
with a slope that is close to the theoretical expectations of $j\propto M^{2/3}$ in a CDM universe 
(see $\S$~\ref{theoryback}).
However, at higher baryon masses, the relation flattens. 
The flattening is mainly driven by galaxy merger activity, which is seen from the 
relation $j_{\rm stars}$ and $j_{\rm bar}$ with stellar mass for galaxies that have undergone different numbers of galaxy mergers 
(Fig.~\ref{JsJbEvoMergers}).
This will be discussed in detail in an upcoming paper (Lagos et al. in prep.). 
We find that the scatter in the $j_{\rm bar}$-$M_{\rm bar}$ relation is well correlated with $V_{\rm rot}/\sigma_{\rm stars}$, $f_{\rm gas,neutral}$, 
$\rm (u^*-r^*)$ and $\rm \langle\rm age_{\rm stars}\rangle$.
We did not find any clear correlation between the positions of galaxies
in the $j_{\rm bar}-M_{\rm bar}$ plane and the
stellar concentration, $r_{90}/r_{50}$, or the
central surface density of stars, and thus we do not show them here.
The middle panels of Fig.~\ref{JBAREAGLE} show the $j_{\rm bar}-M_{\rm bar}$ relation with 
$j_{\rm bar}$ measured within $5\times r_{50}$, for galaxies
with $M_{\rm stars}>10^9\,\rm M_{\odot}$ at $z=0$ in \eagle.
We again find here that the trends seen in Fig.~\ref{JBAREAGLE}  are preserved even if we measure 
$j$ out to large radii. 

The bottom panels of Fig.~\ref{JBAREAGLE} show the $j_{\rm stars}-M_{\rm stars}$ relation with 
$j_{\rm stars}$ measured within $5\times r_{50}$, for galaxies 
with $M_{\rm stars}>10^9\,\rm M_{\odot}$ at $z=0$ in \eagle. 
We colour the plane by the median $\lambda_{\rm R}(5\,r_{50})$ (left panel), 
$f_{\rm gas,neutral}$ (middle left panel), (u*-r*) colour (middle right panel) and mass-weighted stellar age (right panel). 
Here we see that the trends analysed in $\S$~\ref{jz0sec} are also found when we perform the study 
out to large radii. The main difference with Fig.~\ref{JMEAGLE} is that the trend with 
$f_{\rm gas,neutral}$ is stronger when we measure $j_{\rm stars}$ within $5\times r_{50}$.

\end{document}